\documentclass[12pt,english,english,12pt,onecolumn, draftcls]{IEEEtran}
\usepackage[T1]{fontenc}
\usepackage[latin9]{inputenc}
\usepackage{amsthm}
\usepackage{amsmath}
\usepackage{amssymb}
\usepackage{graphicx}
\PassOptionsToPackage{version=3}{mhchem}
\usepackage{mhchem}

\makeatletter
\theoremstyle{plain}
\newtheorem{thm}{\protect\theoremname}
\theoremstyle{definition}
\newtheorem{defn}[thm]{\protect\definitionname}
\theoremstyle{plain}
\newtheorem{prop}[thm]{\protect\propositionname}
\theoremstyle{remark}
\newtheorem{rem}[thm]{\protect\remarkname}

\ifCLASSINFOpdf
\else
\fi

\theoremstyle{plain}
\newtheorem{propo}{Proposition}
\renewenvironment{prop}{\begin{propo}}{\end{propo}}
\theoremstyle{remark}
\newtheorem{rema}{Remark}
\renewenvironment{rem}{\begin{rema}}{\end{rema}}

\makeatother

\usepackage{babel}
\providecommand{\definitionname}{Definition}
\providecommand{\propositionname}{Proposition}
\providecommand{\remarkname}{Remark}
\providecommand{\theoremname}{Theorem}

\begin{document}

\title{Information Embedding on Actions}

\author{Behzad Ahmadi,~\IEEEmembership{Student Member,~IEEE,} Himanshu
Asnani,~\IEEEmembership{Student Member,~IEEE,} Osvaldo Simeone,~\IEEEmembership{Member,~IEEE},
and Haim H. Permuter,~\IEEEmembership{Member,~IEEE}%
\thanks{B. Ahmadi and O. Simeone are with the Center for Wireless Communications
and Signal Processing Research (CWCSPR), ECE Department, New Jersey
Institute of Technology (NJIT), Newark, NJ 07102, USA (email: behzad.ahmadi@njit.edu,
osvaldo.simeone@njit.edu). H. Asnani is with Information Systems Lab
(ISL), Department of Electrical Engineering, Stanford University,
Stanford, CA 94305, USA (email: asnani@stanford.edu). H. H. Permuter
is with the Department of Electrical and Computer Engineering, Ben-Gurion
University of the Negev, Beer-Sheva 84105, Israel (e-mail: haimp@bgu.ac.il)%
}%
\thanks{The work of O. Simeone was supported in part by the U.S. National
Science Foundation under Grant No. 0914899. H. Asnani was supported
by Scott A. and Geraldine D. Macomber Stanford Graduate Fellowship.
H. H. Permuter was supported in part by the Marie Curie Reintegration
fellowship.%
}}
\maketitle
\begin{abstract}
The problem of optimal actuation for channel and source coding was
recently formulated and solved in a number of relevant scenarios.
In this class of models, actions are taken at encoders or decoders,
either to acquire side information in an efficient way or to control
or probe effectively the channel state. In this paper, the problem
of embedding information on the actions is studied for both the source
and the channel coding set-ups. In both cases, a decoder is present
that observes only a function of the actions taken by an encoder or
a decoder of an action-dependent point-to-point link. For the source
coding model, this decoder wishes to reconstruct a lossy version of
the source being transmitted over the point-to-point link, while for
the channel coding problem the decoder wishes to retrieve a portion
of the message conveyed over the link.

For the problem of source coding with actions taken at the decoder,
a single letter characterization of the set of all achievable tuples
of rate, distortions at the two decoders and action cost is derived,
under the assumption that the mentioned decoder observes a function
of the actions non-causally, strictly causally or causally. A special
case of the problem in which the actions are taken by the encoder
is also solved. A single-letter characterization of the achievable
capacity-cost region is then obtained for the channel coding set-up
with actions. Examples are provided that shed light into the effect
of information embedding on the actions for the action-dependent source
and channel coding problems. 
\end{abstract}
\begin{keywords} Action-dependent source coding, action-dependent
channel coding, block Markov decoding, cribbing, forward encoding,
information embedding, source-channel separation, side information,
side information vending machine. \end{keywords}

\section{Introduction}

The recent works \cite{Permuter,Weissman} study the problem of \emph{optimal
actuation for source and channel coding} for resource-constrained
systems. Specifically, in \cite{Permuter}, an extension of the Wyner-Ziv
source coding problem is considered in which the decoder or the encoder
can take actions that affect the quality of the side information available
at the decoder's side. When the actions are taken by the decoder,
the latter operates in two stages. In the first stage, based on the
message received from the encoder, the decoder selects cost-constrained
actions $A$ that affect the measurement of the side information $Y$.
This effect is modelled by a channel $p_{Y|X,A}(y|x,a)$, where $X$
represents the source available at the encoder. In the second stage,
the decoder produces an estimate of source $X$ based on the side
information $Y$ as in the standard Wyner-Ziv problem (see, e.g.,
\cite{Elgammal}). A similar formulation also applies when the actions
are taken at the encoder's side. This model can account, as an example,
for computer networks in which the acquisition of side information
from remote data bases is costly in terms of system resources and
thus should be done efficiently. We refer to this class of problems
as having \emph{actions for} \emph{side information acquisition}.

In \cite{Weissman}, a related channel coding problem is studied in
which the encoder in a point-to-point channel can take actions to
affect the state of a channel. The encoder operates in two stages.
In the first stage, based on the message to be conveyed to the decoder,
cost-constrained actions $A$ are selected by the encoder that affect
the channel state $S$ of the channel $p_{Y|X,S}(y|x,s)$ used for
communication to the decoder in the second stage. In the second stage,
the channel $p_{Y|X,S}(y|x,s)$ is used in a standard way based on
the available information about the state $S$ (which can be non-causal
or causal, see, e.g., \cite{Elgammal}). We refer to this problem
as having \emph{actions for channel state control}. As shown in \cite{probing},
this model can be used to account for an encoder that in the first
stage \emph{probes} the channel to acquire state information.

\subsection{Information Embedding on Actions\label{sub:Information-Embedding-on}}

As discussed above, \emph{optimal actuation for channel and source
coding}, as proposed in\emph{ }\cite{Permuter,Weissman}, prescribes
the selection of the actions $A$ towards the goal of improving the
performance of the resource-constrained communication link between
encoder and decoder. This can be done by acquiring side information
in an efficient way for source coding problems, and by controlling
or probing effectively the channel state for channel coding problems.

This work starts from the observations that the actions $A$ often
entail the use of physical resources for communication within the
system encompassing the link under study. For instance, acquiring
information from a data base requires the receiver to exchange control
signals with a server, and probing the congestion state of a network
(modelled as a channel) requires transmission of training packets
to the closest router. In all these cases, the {}``recipient'' of
the actions, e.g., the server or a router in the examples above, may
request to obtain partial information about the source or message
being communicated on the link. To illustrate this point, the server
in the data base application might need to acquire some explicit information
about the file being transmitted in the link before granting access
to the server. Similarly, the router might need to obtain the header
of the packet (message) that the transmitter intends to deliver to
the end receiver.

\begin{figure}[h!]
\centering \includegraphics[bb=67bp 490bp 519bp 665bp,clip,scale=0.65]{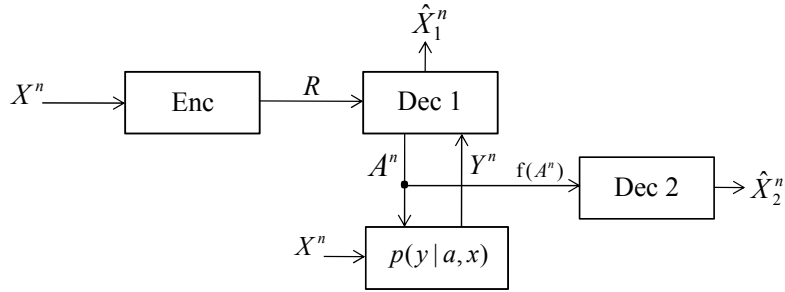}
\caption{Source coding with decoder-side actions for information acquisition
and with information embedding on actions. A function of the actions
$\textrm{f}(A^{n})=(\textrm{f}(A_{1}),...,\textrm{f}(A_{n}))$ is
observed in full ({}``non-causally'') by Decoder 2 before decoding.
See Fig. \ref{fig:fig2} and Fig. \ref{fig:fig3} for the corresponding
models with strictly causal and causal observation of the actions
at Decoder 2, respectively.}

\label{fig:fig1} 
\end{figure}

\begin{figure}[h!]
\centering \includegraphics[bb=69bp 466bp 404bp 628bp,scale=0.65]{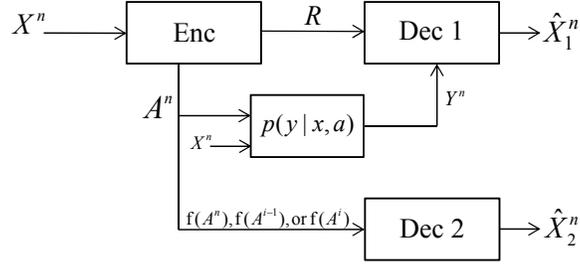}
\caption{Source coding with encoder-side actions for information acquisition
and with information embedding on actions.}

\label{fig:fig4} 
\end{figure}

\begin{figure}[h!]
\centering \includegraphics[bb=40bp 575bp 575bp 750bp,clip,scale=0.65]{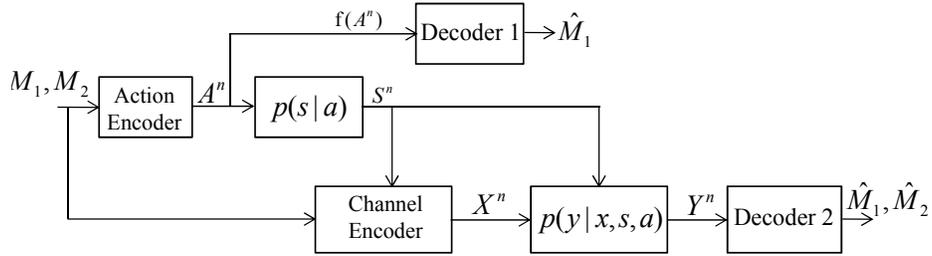}
\caption{Channel coding with actions for channel state control and with information
embedding on actions.}

\label{fig:fig6} 
\end{figure}

In the scenarios discussed above, the action $A$ thus serves a double
purpose: on the one hand, it should be designed to improve the performance
of the communication link at hand as in \cite{Permuter,Weissman,probing},
and, on the other, it should provide explicit information about source
or message for a separate decoder (the server or router in the examples
above). A relevant question thus is: \emph{How much information can
be embedded in the actions $A$ without affecting the performance
of the link}? Or, to turn the question around, \emph{what is the performance
loss for the link as a function of the amount of information that
is encoded in the actions }$A$? This work aims at answering these
questions for both the source and channel coding scenarios discussed
above (see Fig. \ref{fig:fig1}, Fig. \ref{fig:fig4}, Fig. \ref{fig:fig6}
and Sec. \ref{sub:Contributions-and-Paper}).

\subsection{Related Work}

The interplay between communication and actuation, or control, is
recognized to arise at different levels. As mentioned, the main theme
in the papers \cite{Permuter,Weissman,probing} is {}``\emph{control
for communication}'': in \cite{Permuter,Weissman,probing}, actuation
is instrumental in improving the performance of a resource-constrained
communication system. Extensions of this research direction include
models with additional design constraints \cite{Kittichokechai,Asnani real time},
with adaptive actions \cite{Chiru}, with memory \cite{asnani feedback,Asnani real time}
and with multiple terminals \cite{Weissman_multi}\cite{chiru behzad conf}.
Somewhat related, but distinct, is the line of work including \cite{control oriented}-\cite{control oriented 2},
in which control-theoretic tools are leveraged to design effective
communication schemes. An altogether different theme is instead central
in work such as \cite{control over comm 0,control over comm 1} that
can be referred to as {}``\emph{communication for control}''. In
fact, in a reversed way, in \cite{control over comm 0,control over comm 1}
(and references therein), communication is instrumental in carrying
out control tasks such as stabilization of a plant. For instance,
\cite{control over comm 1} shows that an implicit message communicated
between two controllers can greatly improve the performance of the
control task.

The idea of embedding information in the actions is related to the
classical problem of information hiding (see, e.g., \cite{Moulin}
and references therein). In information hiding, a message is embedded
in a host data under distortion constraints. The message is then retrieved
by a decoder that observes the host signal through a noisy channel.
Note that the (host) signal onto which the message is embedded is
a given process. Instead, in the set-up of information embedding on
actions considered here, the (action) signal on which information
is embedded is designed to optimize the given communication task.

The set-up at hand is also related to the source coding model of \cite{Weissman_crib},
in which an encoder communicates to two decoders and one of the decoders
is able to observe the source estimate produced by the other. For
its duality with the classical channel coding model studied in \cite{Willems},
the operation of the first decoder was referred to in \cite{Weissman_crib}
as \emph{cribbing}. Although the problem of interest here (in the
source coding part) is significantly different, in that the recipient
of the embedded information is a decoder {}``cribbing'' the actions
and not the estimates of another decoder, the solutions of the two
problems turn out to be related, as it will be discussed.

\subsection{Contributions and Paper Organization\label{sub:Contributions-and-Paper}}

The main contributions of this paper are as follows. 
\begin{itemize}
\item \textbf{Decoder-side actions for side information acquisition:} We
first consider the model in Fig. \ref{fig:fig1}, in which the problem
of source coding with actions taken at the decoder (Decoder 1) \cite{Permuter}
is generalized by including an additional decoder (Decoder 2). Decoder
2 is the recipient of a function of the action sequence and is interested
in reconstructing a lossy version of the source measured at the encoder.
A single-letter characterization of the set of all achievable tuples
of rate, distortions at the two decoders and action cost is derived
in Sec. \ref{sec:Decoder-Side-Actions-for} under the assumption that
Decoder 2 observes a function of the actions non-causally (Sec. \ref{sub:RD_NCC}),
strictly causally (Sec. \ref{sub:Strictly-Causal-Embedding}) or causally
(Sec. \ref{sub:Causal-Embedding-on}). An example is provided to shed
light into the effect of information embedding on actions in Sec.
\ref{sec:Examples}; 
\item \textbf{Encoder-side actions for side information acquisition: }We
then consider the set-up in Fig. \ref{fig:fig4}, in which an additional
decoder observing the actions is added to the problem of source coding
with actions taken at the encoder \cite{Permuter}. Sec. \ref{sec:Encoder-Side-Actions-for}
derives the achievable rate-distortion-cost region in the special
case in which the channel $p_{Y|X,A}(y|x,a)$ with source and action
($X,A)$ as inputs and side information $Y$ as output is such that
\emph{$Y$} is a deterministic function of $A$; 
\item \textbf{Actions for channel control and probing:} Finally, we consider
the impact of information embedding on actions for channel control
by studying the set-up in Fig. \ref{fig:fig6}, which generalizes
\cite{Weissman}. Specifically, a decoder (Decoder 1) is added to
the model in \cite{Weissman}, that observes a function of the actions
taken by the encoder and wishes to decode part of the message that
is intended for the channel decoder (Decoder 2). A single-letter characterization
of the achievable capacity-cost region is obtained in Sec. \ref{sec:Actions-for-Channel}.
Finally, the special case of actions for channel probing \cite{probing}
is elaborated on with an example in Sec. \ref{sub:Probing-Capacity}. 
\end{itemize}

\section{Decoder-Side Actions for Side Information Acquisition\label{sec:Decoder-Side-Actions-for}}

In this section, we first describe the system model for the set-up
illustrated in Fig. \ref{fig:fig1}, Fig. \ref{fig:fig2} and Fig.
\ref{fig:fig3} of source coding with decoder-side actions. Then,
a single letter characterization of the set of all achievable tuples
of rate, distortions at the two decoders and action cost is derived
under the assumption that Decoder 2 observes the actions fully (non-causally)
in Sec. \ref{sub:RD_NCC}, strictly causally in Sec. \ref{sub:Strictly-Causal-Embedding}
and causally in Sec. \ref{sub:Causal-Embedding-on}. An example is
provided in Sec. \ref{sec:Examples}.

\subsection{System Model\label{sub:System-Model_NCC}}

We present here the problem corresponding to full observation of a
function of the actions as per Fig. \ref{fig:fig1}. We refer to this
model as having non-causal action observation. The changes necessary
to account for causal or strictly causal as illustrated in Fig. \ref{fig:fig2}
and Fig. \ref{fig:fig3} will be discussed in the appropriate sections
later. It is remarked that this definition does not entail any non-causal
operation, but only a larger estimation delay for Decoder 2 as compared
to the causal cases in Fig. \ref{fig:fig2} and Fig. \ref{fig:fig3}.
The model is defined by the probability mass functions (pmfs) $p_{X}(x)$
and $p_{Y|AX}(y|a,x)$, by the function $\textrm{f}\textrm{: }\mathcal{A}\rightarrow\mathcal{B}$,
and by discrete alphabets $\mathcal{X},\mathcal{Y},\mathcal{A},\mathcal{B},\mathcal{\hat{X}}_{1},\mathcal{\hat{X}}_{2},$
as follows. The source sequence $X^{n}$ is such that $X_{i}\in\mathcal{X}$
for $i\in[1,n]$ is independent and identically distributed (i.i.d.)
with pmf $p_{X}(x)$. The Encoder measures sequence $X^{n}$ and encodes
it in a message $M$ of $nR$ bits, which is delivered to Decoder
1. Decoder 1 receives message $M$ and selects an action sequence
$A^{n},$ where $A^{n}\in\mathcal{A}^{n}.$ The action sequence affects
the quality of the measurement $Y^{n}$ of sequence $X^{n}$ obtained
at the Decoder 1. Specifically, given $A^{n}=a^{n}$ and $X^{n}=x^{n}$,
the sequence $Y^{n}$ is distributed as $p(y^{n}|a^{n},x^{n})=\prod_{i=1}^{n}p_{Y|A,X}(y_{i}|a_{i},x_{i})$.
The cost of the action sequence is defined by a cost function $\Lambda$:
$\mathcal{A\rightarrow}[0,\Lambda_{\max}]$ with $0\leq\Lambda_{\max}<\infty,$
as $\Lambda(a^{n})=\sum_{i=1}^{n}\Lambda(a_{i})$. The estimated sequence
$\hat{X}_{1}^{n}\in\mathcal{\hat{X}}_{1}^{n}$ is then obtained as
a function of $M$ and $Y^{n}$. Decoder 2 observes a function of
the action sequence $A^{n}$, thus obtaining $\textrm{f}(A^{n})=(\textrm{f}(A_{1}),...,\textrm{f}(A_{n}))\in\mathcal{B}^{n}$.
Based on $\textrm{f}(A^{n})$, Decoder 2 obtains an estimate $\hat{X}_{2}^{n}\in\mathcal{\hat{X}}_{2}^{n}$
within given distortion requirements. The estimated sequences $\hat{X}_{j}^{n}$
for $j=1,2$ must satisfy distortion constraints defined by functions
$d_{j}(x,\hat{x}_{j})$: $\mathcal{X}\times\mathcal{\hat{X}}_{j}\rightarrow[0,D_{j,\max}]$
with $0\leq D_{j,\max}<\infty$ for $j=1,2,$ respectively. A formal
description of the operations at encoder and decoder follows. 
\begin{defn}
\label{def_cascade}An $(n,R,D_{1},D_{2},\Gamma)$ code for the set-up
of Fig. \ref{fig:fig1} consists of a source encoder 
\begin{equation}
\mathrm{h}^{(e)}\text{: }\mathcal{X}^{n}\rightarrow[1,2^{nR}],\label{encoder1}
\end{equation}
 which maps the sequence $X^{n}$ into a message $M;$ an {}``action\textquotedblright{}\ function
\begin{equation}
\mathrm{h}^{(a)}\text{: }[1,2^{nR}]\rightarrow\mathcal{A}^{n},\label{action_fun}
\end{equation}
 which maps the message $M$ into an action sequence $A^{n};$ two
decoders, namely 
\begin{equation}
\mathrm{h}_{1}^{(d)}\text{: }[1,2^{nR}]\times\mathcal{Y}^{n}\rightarrow\mathcal{\hat{X}}_{1}^{n},\label{decoder1}
\end{equation}
 which maps the message $M$ and the measured sequence $Y^{n}$ into
the estimated sequence $\hat{X}_{1}^{n};$ 
\begin{equation}
\mathrm{h}_{2}^{(d)}\text{: }\mathcal{B}^{n}\rightarrow\mathcal{\hat{X}}_{2}^{n},\label{decoder2}
\end{equation}
 which maps the observed sequence $\textrm{f}(A^{n})$ into the the
estimated sequence $\hat{X}_{2}^{n};$ such that the action cost constraint
$\Gamma$ and distortion constraints $D_{j}$ for $j=1,2$ are satisfied,
i.e., 
\begin{align}
\frac{1}{n}\underset{i=1}{\overset{n}{\sum}}\mathrm{E}\left[\Lambda(A_{i})\right] & \leq\Gamma\label{action cost}\\
\text{ and }\frac{1}{n}\underset{i=1}{\overset{n}{\sum}}\mathrm{E}\left[d_{j}(X_{ji},\hat{X}_{ji})\right] & \leq D_{j}\text{ for }j=1,2.\label{dist const}
\end{align}

\end{defn}

\begin{defn}
\label{def_ach}Given a distortion-cost tuple $(D_{1},D_{2},\Gamma)$,
a rate $R$ is said to be achievable if, for any $\epsilon>0$, and
sufficiently large $n$, there exists a $(n,R,D_{1}+\epsilon,D_{2}+\epsilon,\Gamma+\epsilon)$
code. 
\end{defn}

\begin{defn}
\label{def_reg}The \textit{rate-distortion-cost function }$R(D_{1},D_{2},\Gamma)$
is defined as $R(D_{1},D_{2},\Gamma)=\inf\{R:\textrm{the tuple \ensuremath{(R,D_{1},D_{2},\Gamma)}}\textrm{is achievable}\}$. 
\end{defn}
In the rest of this section, for simplicity of notation, we drop the
subscripts from the definition of the pmfs, thus identifying a pmf
by its argument.

\subsection{Non-Causal Action Observation \label{sub:RD_NCC}}

In this section, a single-letter characterization of the rate-distortion
region is derived for the set-up in Fig. \ref{fig:fig1} in which
Decoder 1 observes the entire sequence $\textrm{f}^{n}(A^{n})$ prior
to decoding. 
\begin{prop}
\label{prop:RD_NCC}The rate-distortion-cost function $\mathcal{\mbox{\ensuremath{R(D_{1},D_{2},\Gamma)}}}$
for the source coding problem with decoder-side actions and non-causal
observation of the actions at Decoder 2 illustrated in Fig. \ref{fig:fig1}
is given by 
\begin{eqnarray}
R(D_{1},D_{2},\Gamma)=\underset{p(\hat{x}_{2},a,u|x),\textrm{ }\ce{g}(U,Y)}{\min}I(X;\hat{X}_{2},A)+I(X;U|\hat{X}_{2},A,Y),\label{eq:R_NCC}
\end{eqnarray}
 where the mutual information is evaluated with respect to the joint
pmf 
\begin{align}
p(x,y,a,\hat{x}_{2},u)=p(x)p(\hat{x}_{2},a,u|x)p(y|x,a) & ,\label{eq:joint}
\end{align}
 for some pmf $p(\hat{x}_{2},a,u|x)$ such that the inequalities\begin{subequations}\label{eqn: action_cascade_const}
\begin{eqnarray}
\ce{E}[d_{j}(X,\hat{X}_{j})] & \leq & D_{j},\textrm{ for \ensuremath{j=1,2,}}\label{eq:dist}\\
\ce{E}[\Lambda(A)] & \leq & \Gamma,\label{eq:action_bound}\\
\ce{and}\textrm{ }I(X;\hat{X}_{2},\ce{f}(A)) & \leq & H(\ce{f}(A))\label{eq:req_NCC}
\end{eqnarray}
 \end{subequations}are satisfied for $\hat{X}_{1}=\ce{g}(U,Y)$ for
some function $\ce{g}\textrm{: }\mathcal{U}\times\mathcal{Y}\rightarrow\hat{\mathcal{X}}_{1}$.
Finally, $U$ is an auxiliary random variable whose alphabet cardinality
can be constrained as $|\mathcal{U}|\leq|\mathcal{X}||\mathcal{\mathcal{\hat{X}}}_{2}||\mathcal{A}|+1$
without loss of optimality. 
\end{prop}
At an intuitive level, in (\ref{eq:R_NCC}), the term $I(X;\hat{X}_{2},A)$
accounts for the rate needed to instruct Decoder 1 about the actions
$A$ to be taken for the acquisition of the side information $Y$,
which are selected on the basis of the source $X$, and, at the same
time, to communicate the reconstruction $\hat{X}_{2}$ to Decoder
2. The additional rate $I(X;U|\hat{X}_{2},A,Y)$ is instead required
to refine the description of the source $X$ provided via $(\hat{X}_{2},A)$
using an auxiliary codebook $U$ for Decoder 1. Note that this rate
is conditioned on the side information $Y$, thanks to the rate saving
obtained through Wyner-Ziv binning. The condition (\ref{eq:req_NCC})
ensures that, based on the observation of $\ce{f}(A)$, Decoder 2
is able to reconstruct $\hat{X}_{2}$. The details of achievability
follow as a combination of the techniques proposed in \cite{Permuter}
and \cite{cuff zhao,Weissman_crib}. Below we briefly outline the
main ideas, since the technical details follow from standard arguments.
The proof of the converse is provided in Appendix A.

\emph{Sketch of the achievability proof}: We fix a pmf (\ref{eq:joint})
and define a random variable $B=\textrm{f}(A)$. The joint pmf $p(x,y,a,\hat{x}_{2},u,b)$
of variables $(X,Y,A,\hat{X}_{2},U,B)$ is obtained by multiplying
the right-hand side of (\ref{eq:joint}) by the term%
\footnote{The notation $\mathbf{1}_{\{S\}}$ is used for the indicator function
of the event $S$.%
} $\mathbf{1}_{\{b=\textrm{f}(a)\}}$. In the scheme at hand, the Encoder
first maps sequence $X^{n}$ into a sequence $\hat{X}_{2}^{n}\in\mathcal{\hat{X}}_{2}^{n}$
using the joint typicality criterion with respect to the joint pmf
$p(x,\hat{x}_{2})$. This mapping requires a codebook of rate $I(X;\hat{X}_{2})$
(see, e.g., \cite[pp. 62-63]{Elgammal})$.$ Given the sequence $\hat{X}_{2}^{n}$,
the sequence $X^{n}$ is further mapped into a sequence $B^{n}\in\mathcal{B}^{n}$
using the joint typicality criterion with respect to the joint pmf
$p(x,b|\hat{x}_{2})$ where $B=\textrm{f}(A)$, which requires a codebook
of rate $I(X;\textrm{f}(A)|\hat{X}_{2})$ for each sequence $\hat{X}_{2}^{n}$.
For later reference, we refer to every such codebook as a bin in the
following. Note that we have one bin for every sequence $\hat{X}_{2}^{n}$.
For each pair $(\hat{X}_{2}^{n},B^{n})$, the sequence $X^{n}$ is
mapped into an action sequence $A^{n}$ using joint typicality with
respect to the joint pmf $p(x,a|\hat{x}_{2},b)$, which requires a
codebook of rate $I(X;A|\hat{X}_{2},\textrm{f}(A))$. Note that, by
construction, we have that $B^{n}=\textrm{f}(A^{n})$ for each generated
$A^{n}$. Finally, the source sequence $X^{n}$ is mapped into a sequence
$U^{n}$ using the joint typicality criterion with respect to the
joint pmf $p(x,u|\hat{x}_{2},a)$, which requires a codebook of rate
$I(X;U|\hat{X}_{2},A)$ for each pair $(\hat{X}_{2}^{n},A^{n})$.

The indices of codewords $\hat{X}_{2}^{n}$, $B^{n}$ and $A^{n}$
are sent to Decoder 1, along with the index for the codeword $U^{n}$.
For the latter, by leveraging the side information $Y^{n}$ available
at Decoder 1, the rate can be reduced to $I(X;U|\hat{X}_{2},A,Y)$
by the Wyner-Ziv theorem \cite[p. 280]{Elgammal}. Decoder 2 estimates
the sequence $\hat{X}_{2}^{n}$ from the observed sequence $\textrm{f}(A^{n})$
as follows: if there is only one bin containing the observed sequence
$\textrm{f}(A^{n})$, then $\hat{X}_{2}^{n}$ equals the sequence
corresponding to such bin (recall that each bin corresponds to one
sequence $\hat{X}_{2}^{n}$). Otherwise, an error is decoded. To obtain
a vanishing probability of error, the sequence $\textrm{f}^{n}(A^{n})$
should thus not lie within more than one bin with high probability.
The probability of the latter event can be upper bounded by $2^{n(I(X;\hat{X}_{2},\textrm{f}(A))-H(\textrm{f}(A))}$
since each sequence $B^{n}$ is generated with probability approximately
$2^{-nH(\textrm{f}(A))}$ and there are $2^{nI(X;\hat{X}_{2},\textrm{f}(A))}$
sequences $B^{n}$ \cite{Weissman_crib}. Therefore, as long as $I(X;\hat{X}_{2},\textrm{f}(A))\leq H(\textrm{f}(A))$,
Decoder 1 is able to infer the conveyed bin index with high probability.
Finally, Decoder 1 produces the estimate $\hat{X}_{1}^{n}$ through
a symbol-by-symbol function as $\hat{X}_{1i}=\textrm{g}(U_{i},Y_{i})$
for $i\in[1,n].$\hfill{}$\Box$ 
\begin{rem}
\label{rem:Assume-that-the}Assume that the action $A_{i}$ is allowed
to be a function, not only of the message $M$ as per (\ref{action_fun}),
but also of the previous values of the side information $Y^{i-1}$,
which we refer to as adaptive actions. Then, the rate-distortion-cost
function derived in Proposition \ref{prop:RD_SCC} can generally be
improved. This can be seen by considering the case in which $R=0$.
In this case, if the actions were selected as per (\ref{action_fun}),
then the distortion at Decoder 2 would be forced to be maximal, i.e.,
$D_{2}=D_{2,\mathrm{max}}$, since the actions $A$ cannot depend
in any way on the source $X$. Instead, by selecting $A$ as a function
of the previously observed values of $Y$, Decoder 1 can provide Decoder
2 with information about $X$, thus decreasing the distortion $D_{2}$.
It is noted that the usefulness of adaptive actions in this setting
contrasts with the known fact that, in the absence of Decoder 2, adaptive
actions do not decrease the rate-distortion function \cite{Chiru}. 
\end{rem}

\subsection{Strictly Causal Action Observation \label{sub:Strictly-Causal-Embedding}}

\begin{figure}[h!]
\centering \includegraphics[bb=67bp 490bp 519bp 665bp,clip,scale=0.65]{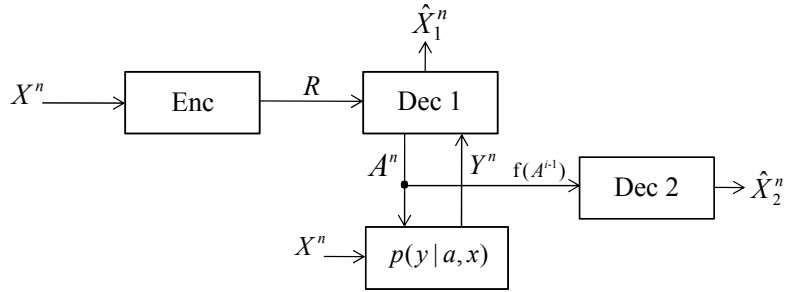}
\caption{Source coding with decoder-side actions for information acquisition
and with information embedding on actions. At time $i$, Decoder 2
has available the samples $\textrm{f}(A^{i-1})=(\textrm{f}(A_{1}),...,\textrm{f}(A_{i-1}))$
in a strictly causal fashion.}

\label{fig:fig2} 
\end{figure}

The system model for the set-up in Fig. \ref{fig:fig2}, is similar
to the one described in Sec. \ref{sub:System-Model_NCC} with the
only difference the decoding function for Decoder 2 a time $i$ is
given as 
\begin{equation}
\mathrm{h}_{2i}^{(d)}\text{: }\mathcal{B}^{i-1}\rightarrow\mathcal{\hat{X}}_{2},\label{decoder2_SCC}
\end{equation}
 which maps the strictly causally observed sequence $\textrm{f}(A^{i-1})=(\textrm{f}(A_{1}),...,\textrm{f}(A_{i-1}))$
into the $i$th estimated symbol $\hat{X}_{2i}$. 

\begin{prop}
\label{prop:RD_SCC}The rate-distortion-cost function $\mathcal{\mbox{\ensuremath{R(D_{1},D_{2},\Gamma)}}}$
for the source coding problem with decoder-side actions and strictly
causal observation of the actions at Decoder 1 as illustrated in Fig.
\ref{fig:fig2} is given by 
\begin{eqnarray}
R(D_{1},D_{2},\Gamma)=\underset{p(\hat{x}_{2},a,u|x),\textrm{ }\ce{g}(U,Y)}{\min}I(X;\hat{X}_{2},A)+I(X;U|\hat{X}_{2},A,Y),\label{eq:R_SCC}
\end{eqnarray}
 where the mutual information is evaluated with respect to the joint
pmf 
\begin{align}
p(x,y,a,\hat{x}_{2},u)=p(x)p(\hat{x}_{2},a,u|x)p(y|x,a) & ,\label{eq:joint_SCC}
\end{align}
 for some pmf $p(\hat{x}_{2},a,u|x)$ such that the inequalities\begin{subequations}\label{eqn: action_cascade_const-2}
\begin{eqnarray}
\ce{E}[d_{j}(X,\hat{X}_{j})] & \leq & D_{j},\textrm{ for \ensuremath{j=1,2,}}\label{eq:dist_SCC}\\
\ce{E}[\Lambda(A)] & \leq & \Gamma,\label{eq:action_bound_SCC}\\
\ce{and}\textrm{ }I(X;\hat{X}_{2},\ce{f}(A)) & \leq & H(\ce{f}(A)|\hat{X}_{2})\label{eq:req_SCC}
\end{eqnarray}
 \end{subequations}are satisfied for $\hat{X}_{1}=\ce{g}(U,Y)$ for
some function $\ce{g}\textrm{: }\mathcal{U}\times\mathcal{Y}\rightarrow\hat{\mathcal{X}}_{1}$.
Finally, $U$ is an auxiliary random variable whose alphabet cardinality
can be constrained as $|\mathcal{U}|\leq|\mathcal{X}||\mathcal{\mathcal{\hat{X}}}_{2}||\mathcal{A}|+1$
without loss of optimality. 
\end{prop}
The only difference between the rate-distortion-cost function of Proposition
\ref{prop:RD_NCC} with non-causal action observation with respect
to the case with strictly causal action observation of Proposition
\ref{prop:RD_SCC} is the constraint (\ref{eq:req_SCC}). Recall that
the latter is needed to ensure that Decoder 2 is able to recover the
reconstruction $\hat{X}_{2}$. As detailed below, the strict causality
of the observation of the action at Decoder 2 calls for a block-based
encoding in which the actions carries information about the source
sequence as observed in two different blocks, namely the current block
for Decoder 1 and the future block for Decoder 2. This additional
requirement causes the conditioning on $\hat{X}_{2}$ in (\ref{eq:req_SCC}),
which generally increases the rate (\ref{eq:R_SCC}) with respect
to the counterpart (\ref{eq:R_NCC_enc}) achievable with non-causal
action observation. A sketch of the achievability proof is provided
below and is based on the techniques proposed in \cite{cuff zhao,Weissman_crib}
(see also \cite{Pradhan}). The proof of the converse is provided
in Appendix B.

\emph{Sketch of the achievability proof}: We fix a pmf (\ref{eq:joint_SCC})
and define a random variable $B=\textrm{f}(A)$. The joint pmf $p(x,y,a,\hat{x}_{2},u,b)$
of variables $(X,Y,A,\hat{X}_{2},U,B)$ is obtained by multiplying
the right-hand side of (\ref{eq:joint_SCC}) by the term $1_{\{b=\textrm{f}(a)\}}$.
We use the {}``Forward Encoding'' and {}``Block Markov Decoding''
strategy of \cite{cuff zhao,Weissman_crib} (see also \cite{Pradhan})
and combine it with the coding scheme of \cite{Permuter}. The scheme
operates over multiple blocks and we denote by $X^{n}(l)$ the portion
of the source sequence encoded in block $l$. The sequence $\textrm{f}(A^{n}(l))$
observed during block $l$ is used in block $l+1$ by Decoder 2 due
to the strict causality constraint. To this end, the action sequence
$\textrm{f}(A^{n}(l))$ produced in block $l$ must carry information
about the source sequence $X^{n}(l+1)$ corresponding to the next
block $l+1$. Note that this is possible since encoder knows the entire
sequence $X^{n}$. At the same time, sequence $A^{n}(l)$ should also
perform well as an action sequence to be used by Decoder 1 to estimate
sequence $X^{n}(l)$ for the current block. This is accomplished as
follows; In each block $l$, $2^{nI(X;\hat{X}_{2})}$ codewords $\hat{X}_{2}^{n}\in\mathcal{\hat{X}}_{2}^{n}$
are generated according to the pmf $p(\hat{x}_{2})$. Next, $2^{nI(X;\hat{X}_{2})}$
bins are assigned to each codeword $\hat{X}_{2}^{n}$, where each
bin contains $2^{nI(X;\textrm{f}(A)|\hat{X}_{2})}$ codewords $B^{n}\in\mathcal{B}^{n}$,
generated according to pmf $p(b|\hat{x}_{2}).$ For each pair $(\hat{X}_{2}^{n},B^{n})$,
a codebook of $2^{nI(X;A|\hat{X}_{2},\textrm{f}(A))}$ codewords $A^{n}\in\mathcal{A}^{n}$
is generated according to the joint pmf $p(x,a|\hat{x}_{2},b)$. Finally,
a codebook of $2^{nI(X;U|\hat{X}_{2},A)}$ codewords $U^{n}\in\mathcal{U}^{n}$
is generated according to the joint pmf $p(x,u|\hat{x}_{2},a)$. The
latter codebook is further binned into a codebook of rate $I(X;U|\hat{X}_{2},A,Y)$
to leverage the side information $Y^{n}$ available at Decoder 1 via
the Wyner-Ziv theorem \cite[p. 280]{Elgammal}.

For encoding, in each block $l$, a sequence $\hat{X}_{2}^{n}$ is
selected from the $\hat{X}_{2}-$codebook of block $l$ to be jointly
typical with the source sequence $X^{n}(l)$ in the current block.
Instead, the bin index describes a $\hat{X}_{2}^{n}$ sequence in
the $\hat{X}_{2}-$codebook of block $(l+1)$th that is jointly typical
with the source sequence $X^{n}(l+1)$ of the $(l+1)$th block. Moreover,
given $\hat{X}_{2}^{n}$ and the bin index, a sequence $A^{n}$ is
chosen such that $(A^{n},X^{n}(l))$ are jointly typical. Similarly,
a sequence $U^{n}$ is selected for block $l$ to be jointly typical
with the sequence of $X^{n}(l)$ of block $l$.

Thanks to the observation of the actions, at block $l+1$ Decoder
2 knows the functions $\textrm{f}(A^{n}(l))$, and aims to find the
bin index in which the corresponding codeword $B^{n}$ lies. As shown
in \cite{Weissman_crib}, this is possible with vanishing probability
of error, if $I(X;\hat{X}_{2},\ce{f}(A))\leq H(\ce{f}(A)|\hat{X}_{2})$.
Note that the conditioning in the right-hand side is due to the fact
that the sequences $B^{n}$ are generated conditioned on the sequence
$\hat{X}_{2}^{n}$ representing a compressed version of the source
for the current block $l$. The latter does not bring any information
regarding the desired sequence $X^{n}(l+1)$.\hfill{}$\Box$ 
\begin{rem}
\label{adaptive_action}From the proof of the converse in Appendix
B, it follows, similarly to \cite{Chiru}, that, adaptive actions
(see Remark
\begin{figure}[h!]
\centering \includegraphics[bb=67bp 490bp 519bp 665bp,clip,scale=0.65]{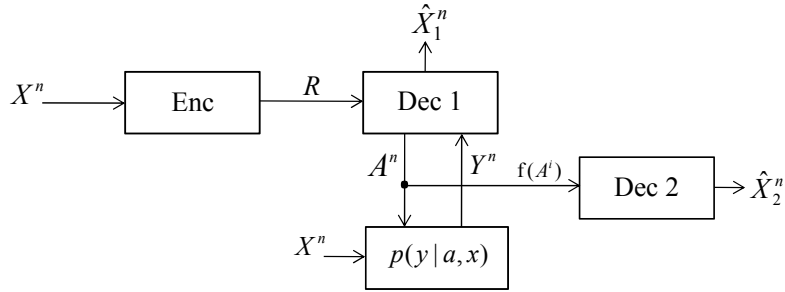}
\caption{Source coding with decoder-side actions for information acquisition
and with information embedding on actions. At time $i$, Decoder 2
has available the samples $\textrm{f}(A^{i})=(\textrm{f}(A_{1}),...,\textrm{f}(A_{i}))$
in a causal fashion.}

\label{fig:fig3} 
\end{figure}
 \ref{rem:Assume-that-the}) do not increase the rate-distortion-cost
function derived in Proposition \ref{prop:RD_SCC}. 
\end{rem}

\subsection{Causal Action Observation \label{sub:Causal-Embedding-on}}

The system model for the set-up in Fig. \ref{fig:fig3}, is similar
to the one described in Sec. \ref{sub:System-Model_NCC} with the
only difference the decoding function for Decoder 2 is 
\begin{equation}
\mathrm{h}_{2i}^{(d)}\text{: }\mathcal{B}^{i}\rightarrow\mathcal{\hat{X}}_{2},\label{decoder2_CC}
\end{equation}
 which maps the causally observed sequence $\textrm{f}(A^{i})=(\textrm{f}(A_{1}),...,\textrm{f}(A_{i}))$
into the $i$th estimated symbol $\hat{X}_{2i}$. 
\begin{prop}
\label{prop:RD_CC}The rate-distortion-cost function $\mathcal{\mbox{\ensuremath{R(D_{1},D_{2},\Gamma)}}}$
for the source coding problem with decoder-side actions and causal
observation of the actions illustrated in Fig. \ref{fig:fig3} is
given by 
\begin{eqnarray}
R(D_{1},D_{2},\Gamma)=\underset{p(v,a,u|x),\textrm{ }\ce{g}_{1}(U,Y),\textrm{ }\ce{g}_{2}(V,\ce{f}(A))}{\min}I(X;V,A)+I(X;U|V,A,Y),\label{eq:R1-1-1}
\end{eqnarray}
 where the mutual information is evaluated with respect to the joint
pmf 
\begin{align}
p(x,y,a,u,v)=p(x)p(v,a,u|x)p(y|x,a) & ,\label{eq:joint-1-1}
\end{align}
 for some pmf $p(v,a,u|x)$ such that the inequalities\begin{subequations}\label{eqn: action_cascade_const-2-1}
\begin{eqnarray}
\ce{E}[d_{j}(X,\hat{X}_{j})] & \leq & D_{j},\textrm{ for \ensuremath{j=1,2,}}\label{eq:dist-1-1}\\
\ce{E}[\Lambda(A)] & \leq & \Gamma,\label{eq:action_bound-1-1}\\
\ce{and}\textrm{ }I(X;V,\ce{f}(A)) & \leq & H(\ce{f}(A)|V)\label{eq:req_SCC-1}
\end{eqnarray}
 \end{subequations}are satisfied for $\hat{X}_{1}=\ce{g}_{1}(U,Y)$
and $\hat{X}_{2}=\ce{g}_{2}(V,\ce{f}(A))$ with some functions $\ce{g}_{1}\textrm{: }\mathcal{U}\times\mathcal{Y}\rightarrow\hat{\mathcal{X}}_{1}$
and $\ce{g}_{2}\textrm{: }\mathcal{V}\times\mathcal{B}\rightarrow\hat{\mathcal{X}}_{2}$,
respectively. Finally, $U$ and $V$ are auxiliary random variables
whose alphabet cardinalities can be constrained as $|\mathcal{U}|\leq|\mathcal{X}||\mathcal{\mathcal{V}}||\mathcal{A}|+1$
and $|\mathcal{V}|\leq|\mathcal{X}|+3$, respectively, without loss
of optimality. 
\end{prop}
The difference between the rate-distortion-cost function above with
causal and strictly causal action observation is given by the fact
that, with causal action observation, Decoder 2 can use the current
value of the function $\ce{f}(A)$ for the estimate of $\hat{X}_{2}$.
This is captured by the fact that the Encoder provides Decoder 2 with
an auxiliary source description $V$, which is then combined with
$\ce{f}(A)$ via a function $\hat{X}_{2}=\ce{g}_{2}(V,\ce{f}(A))$
to obtain $\hat{X}_{2}$. The rate (\ref{eq:R1-1-1}) and the constraint
(\ref{eq:req_SCC-1}) are changed accordingly. The proof of the converse
is given in Appendix B. 
\begin{rem}
As seen in Appendix B, with adaptive actions, the rate-distortion-cost
function derived in Proposition \ref{prop:RD_CC} remains unchanged. 
\end{rem}

\subsection{Binary Example\label{sec:Examples}}

\begin{figure}[h!]
\centering \includegraphics[bb=145bp 380bp 530bp 585bp,clip,scale=0.65]{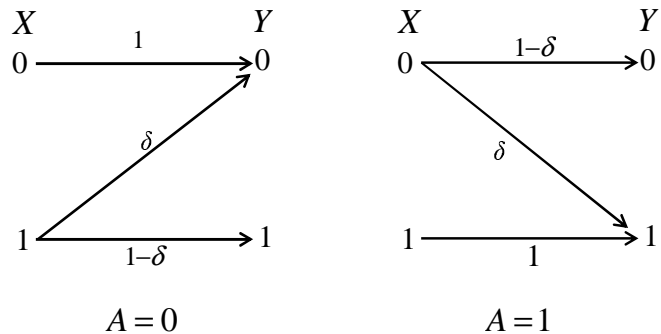}
\caption{The side information channel $p(y|x,a)$ used in the example of Sec.
\ref{sec:Examples}.}

\label{fig:S-channel} 
\end{figure}

In this section, an example is provided to illustrate the effect of
the communication requirements of the additional decoder (Decoder
2) that observes a function of the actions on the system performance.
We assume binary alphabets as ${\cal X}={\cal A}={\cal Y}=\{0,1\}$
and a source distribution $X\sim\textrm{Bern}(\frac{1}{2})$. The
distortion metrics are assumed to be Hamming, i.e., $d_{j}(x,\hat{x}_{j})=0$
if $x=\hat{x}_{j}$ and $d_{j}(x,\hat{x}_{j})=1$ otherwise for $j=1,2$.
Moreover, as shown in Fig. \ref{fig:S-channel}, the side information
$Y$ at Decoder 1 is observed through a Z-channel for $A=0$ or an
S-channel for $A=1$. We assume no cost constraint on the actions
taken by Decoder 1 (which can be enforced by choosing $\Lambda(A)=A$
and $\Gamma=1$), and we set $\textrm{f}(A)=A$. The example extends
that of \cite[Sec. II-D]{Permuter} to a set-up with the additional
Decoder 2. Under the requirement of lossless reconstruction at Decoder
1, i.e., $D_{1}=0$, the rate-distortion-cost function $R(0,D_{2},\Gamma=1)$
with non-causal action observation is obtained from Proposition \ref{prop:RD_NCC}
by setting $U=\hat{X}_{1}=X$, obtaining 
\begin{eqnarray}
R(0,D_{2},1)=\underset{p(\hat{x}_{2},a|x)}{\min}I(X;\hat{X}_{2},A)+H(X|\hat{X}_{2},A,Y),\label{eq:R_NCC_ex}
\end{eqnarray}
 where the minimization is done under the constraints $\ce{E}[d_{2}(X,\hat{X}_{2})]\leq D_{2}$
and $I(X;\hat{X}_{2},A)\leq H(A)$. 

The minimization in (\ref{eq:R_NCC_ex}) can be done over the parameters
$p(a=1,\hat{x}_{2}=0|x=0)\overset{\triangle}{=}\alpha_{1}$, $p(a=1,\hat{x}_{2}=1|x=0)\overset{\triangle}{=}\alpha_{2}$
and $p(a=0,\hat{x}_{2}=1|x=0)\overset{\triangle}{=}\alpha_{3}$ with
$\overset{3}{\underset{i=1}{\sum}}\alpha_{i}\leq1$ and $\alpha_{i}\geq0$
for $i=1,2,3$, since by symmetry, we can set $p(a=0,\hat{x}_{2}=1|x=1)\overset{}{=}\alpha_{1}$,
$p(a=0,\hat{x}_{2}=0|x=1)\overset{}{=}\alpha_{2}$ and $p(a=1,\hat{x}_{2}=0|x=1)\overset{}{=}\alpha_{3}$
without loss of optimality. Explicit expressions can be easily found
and have been optimized numerically.

\begin{figure}[h!]
\centering \includegraphics[bb=35bp 250bp 560bp 675bp,clip,scale=0.55]{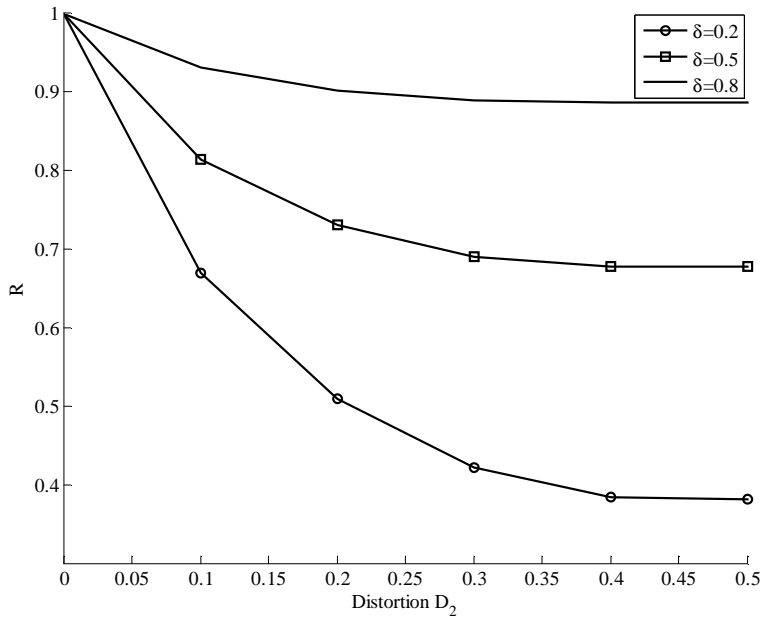}
\caption{Rate-distortion function $R(0,D_{2},1)$ in (\ref{eq:R_NCC_ex}) versus
distortion $D_{2}$ with the side information channel in Fig. \ref{fig:S-channel}
(non-causal side information).}

\label{fig:plot} 
\end{figure}

Fig. \ref{fig:plot} depicts the rate-distortion function versus the
distortion $D_{2}$ of Decoder 2 for values of $\delta=0.2$, $\delta=0.5$
and $\delta=0.8$. It can be seen that if the distortion $D_{2}$
tolerated by Decoder 2 is sufficiently large (e.g., $D_{2}\geq0.4$
for $\delta=0.5$), then the communication requirements of Decoder
2 do not increase the required rate. This can be observed by comparing
the rate $R(0,D_{2},\Gamma)$ with rate $R(0,0.5,\Gamma)$ corresponding
to a distortion level $D_{2}=0.5$, which requires no communication
to Decoder 2. The smallest distortion $D_{2}$ that does not affect
the rate can be found as $D_{2}=\alpha_{2,opt}+\alpha_{3,opt}$, where
$\alpha_{2,opt}$ and $\alpha_{3,opt}$ are the optimal values for
problem (\ref{eq:R_NCC_ex}) with $D_{2}=0.5$ that minimizes $\alpha_{2}+\alpha_{3}$.

\begin{figure}[h!]
\centering\includegraphics[bb=37bp 235bp 570bp 650bp,clip,scale=0.55]{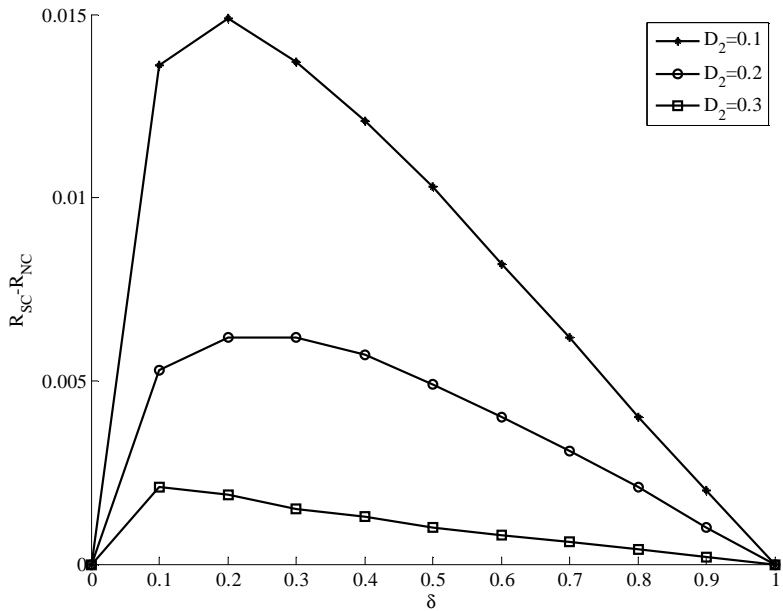}
\caption{Difference between the rate-distortion function (\ref{eq:R_NCC_ex})
with non-causal (NC) and strictly causal (SC) action observation versus
$\delta$ for values of distortion $D_{2}=0.1,$ $0.2$ and $0.3$.}

\label{fig:plot_ncvsc} 
\end{figure}

We now compare the performance between non-causal action observation,
as considered above, and strictly causal action observation. The performance
in the latter case can be obtained from Proposition \ref{prop:RD_SCC}
and leads to (\ref{eq:R_NCC_ex}) with the more restrictive constraint
(\ref{eq:req_SCC}). Fig. \ref{fig:plot_ncvsc} plots the difference
between rate-distortion function (\ref{eq:R_NCC_ex}) for the case
of non-causal and strictly causal action observation versus $\delta$
for three values of distortion, namely $D_{2}=0.1,0.2,0.3.$ As shown,
irrespective of the value of distortion $D_{2}$, for values of $\delta=0$
and $\delta=1$, the performance with non-causal action observation
is equal to that with strictly causal observation. This is due to
the facts that: \emph{i}) for $\delta=0$, the side information $Y$
is a noiseless measure of the source sequence $X$ for both $A=0$
and $A=1$ and thus there is no gain in making the actions at Decoder
1 to be dependent of $X$, and thus $\hat{X}_{2}$; \emph{ii}) for
$\delta=1$, the side information $Y$ is independent of the source
sequence $X$ given both $A=0$ and $A=1$, and thus it is without
loss of optimality to choose actions at Decoder 1 to be independent
of $X$ and $\hat{X}_{2}$. We can conclude that for both $\delta=0$
and $\delta=1$, causal action observation, and in fact even selecting
$A$ to be independent of $X$, does not entail any performance loss.
Instead, for values $0<\delta<1$, it is generally advantageous for
Decoder 1 to select actions correlated with the source $X$, and hence
some performance loss is observed with strictly causal action observation
owing to the more restrictive constraint (\ref{eq:req_SCC}). This
reflects the need to cater to both Decoder 1 and Decoder 2 when selecting
actions $A$, which requires description of two different source blocks.
Following similar arguments, it is also noted that, as the communication
requirements for Decoder 2 become more pronounced, i.e., as $D_{2}$
decreases, the difference between the rate-distortion function with
non-causal and strictly-causal action observation increases. The performance
with causal action observation is intermediate between full and strictly
causal observation, and it is not shown here.

\section{Encoder-Side Actions for Side Information Acquisition\label{sec:Encoder-Side-Actions-for}}

In the previous section, the actions controlling the quality and availability
of the side information were taken by the decoder. In this section,
following \cite[Sec. III]{Permuter}, we consider instead scenarios
in which the encoder takes the actions affecting the side information
of Decoder 1, as shown in Fig. \ref{fig:fig4}. Specifically, the
encoder takes actions $A^{n}\in\mathcal{A}^{n}$, thus influencing
the side information available to the Decoder 1 through a discrete
memoryless channel $p(y|x,a)$. Decoder 2 observes the action sequence
to obtain the deterministic function $\textrm{f}(A^{n})=(\textrm{f}(A_{1}),...,\textrm{f}(A_{n}))$,
or the corresponding causal and strictly causal function, which is
used to estimate the source sequence subject to a distortion constraint.

An $(n,R,D_{1},D_{2},\Gamma)$ code is defined similar to the previous
sections with the difference that the action encoder (\ref{action_fun})
maps directly the source sequence $X^{n}$ into the action sequence
$A^{n}$, i.e., 
\begin{equation}
\mathrm{h}^{(a)}\text{: }\mathcal{X}^{n}\rightarrow\mathcal{A}^{n}.\label{action_fun_enc}
\end{equation}
 As discussed in \cite{Permuter}, even in the absence of Decoder
2, the problem at hand is challenging. We thus focus on certain special
cases, first the special case in which the side information channel
$p(y|x,a)$ is such that $Y$ is a deterministic function of $A$,
i.e., $Y=\textrm{f}_{Y}(A)$, and $\textrm{f}(A)=A$. This is solved
in Proposition \ref{prop:RD_E_NCC}, and generalized by the following
remark to the case of all deterministic function $f$ for which $H(\textrm{f}_{Y}(A)|\textrm{f}(A))=0$.
Following Proposition \ref{prop:RD_E_NCC} is Proposition \ref{prop_new},
where the case $H(\textrm{f}(Y)|\textrm{f}_{Y}(A))=0$ is solved. 
\begin{prop}
\label{prop:RD_E_NCC}The rate-distortion-cost function $\mathcal{\mbox{\ensuremath{R(D_{1},D_{2},\Gamma)}}}$
for the source coding problem with encoder-side actions, non-causal,
causal or strictly causal observation of the actions illustrated in
Fig. \ref{fig:fig4} with $\ce{f}(A)=A$ and $Y=\ce{f}_{Y}(A)$ is
given by 
\begin{eqnarray}
R(D_{1},D_{2},\Gamma)=\underset{p(u|x),\textrm{ }p(\hat{x}_{1}|u,x),\textrm{ }p(\hat{x}_{2}|u,x),\textrm{ }p(a)}{\min}\{I(X;\hat{X}_{1},U)-H(\ce{f}_{Y}(A))\}^{+},\label{eq:R_NCC_enc}
\end{eqnarray}
 where the information measures are evaluated with respect to the
joint pmf 
\begin{align}
p(x,u,\hat{x}_{1},\hat{x}_{2},a)=p(x)p(u|x)p(\hat{x}_{1}|u,x)p(\hat{x}_{2}|u,x)p(a) & ,\label{eq:joint_enc}
\end{align}
 for some pmfs $p(u|x),$ \textup{$p(\hat{x}_{1}|u,x),$ $p(\hat{x}_{2}|u,x)$}
and $p(a)$ such that the inequalities \begin{subequations}\label{eqn: action_const_enc}
\begin{eqnarray}
\ce{E}[d_{j}(X,\hat{X}_{j})] & \leq & D_{j},\textrm{ for \ensuremath{j=1,2,}}\\
\ce{E}[\Lambda(A)] & \leq & \Gamma,\\
I(X;U) & \leq & H(\ce{f}_{Y}(A))\label{eq:enc_req1}\\
\ce{and}\textrm{ }I(X;\hat{X}_{2}|U) & \leq & H(A|\ce{f}_{Y}(A))\label{eq:enc_req2}
\end{eqnarray}
 \end{subequations}are satisfied. Finally, $U$ is an auxiliary random
variable whose alphabet cardinality can be constrained as $|\mathcal{U}|\leq|\mathcal{X}||\mathcal{\hat{X}}_{1}||\mathcal{\hat{X}}_{2}|+3$
without loss of optimality. \end{prop}
\begin{rem}
The results above generalizes a number of known single-letter characterizations.
Notably, if $D_{2}=D_{2,\textrm{max}}$, so that the distortion requirements
of Decoder 2 are immaterial to the system performance, the result
reduces to \cite[Theorem 7]{Permuter}. Moreover, in the special case
in which $A=(A_{0},A_{2})$, $Y=A_{0}$, $R=R_{1}$, $|{\mathcal{A}_{0}}|=2^{R_{0}}$,
$|{\mathcal{A}_{2}}|=2^{R_{2}}$, the model coincides with the lossy
Gray-Wyner problem \cite{Gray}%
\footnote{Note that here $2^{R_{0}}$ and $2^{R_{2}}$ are constrained to be
integers.%
}. 
\end{rem}
As detailed in the proof below, Proposition \ref{prop:RD_E_NCC} establishes
the optimality of separate source-channel coding for the set-up in
Fig. \ref{fig:fig4} under the stated conditions. In particular, the
encoder compresses using a standard successive refinement source code
in which $U$ represents the coarse description and $\hat{X}_{1},\hat{X}_{2}$
two independent refinements. The indices of the coarse description
$U$ and of the refined description $\hat{X}_{2}$ are sent on the
degraded (deterministic) broadcast channel with input $A$ and outputs
$(A,\ce{t}(A))$ using superposition coding. Reliable compression
and communication is guaranteed by the two bounds (\ref{eq:enc_req1})-(\ref{eq:enc_req2}).
A further refined description $\hat{X}_{1}$ is produced for Decoder
1, and the corresponding index is sent partly over the mentioned broadcast
channel and partly over the link of rate $R$, leading to the rate
(\ref{eq:R_NCC_enc}). Details of the achievability proof can be found
below, while the proof of the converse is given in Appendix C. 
\begin{rem}
Following the discussion above, specializing Proposition \ref{prop:RD_E_NCC}
to the case $R=0$ shows the optimality of source-channel coding separation
for the lossy transmission of a source over a deterministic degraded
broadcast channel (see \cite[Chapter 14]{Elgammal} for a review of
scenarios in which the optimality of separation holds for lossless
transmission over a broadcast channel). 
\end{rem}
\emph{Sketch of the achievability proof}: As anticipated above, achievability
uses the ideas of a source-channel coding separation, successive refinement
and superposition coding. We only describe the outline, as the rigorous
details can be derived based on standard techniques \cite{Elgammal}.
We start with the case of non-causal action observation at Decoder
2. Note that the deterministic channel with input $A$ and outputs
$A$ (to Decoder 2) and $\ce{f}_{Y}(A)$ (to Decoder 1) is not only
deterministic but also degraded \cite[Chapter 5]{Elgammal}. This
channel is used to send a common source description of rate $\tilde{R}_{1}$
to both the decoders and a refined description of rate $\tilde{R}_{2}$
to Decoder 2 only. To elaborate, fix the pmfs $p(u|x)$, $p(\hat{x}_{1}|u,x)$,
$p(\hat{x}_{2}|u,x)$ and $p(a)$. Generate a codebook of $2^{nI(X;U)}$
sequences $U^{n}$ i.i.d. with the pmf $p(u)$ and, for each $U^{n}$
sequence, generate a codebook of $2^{nI(X;\hat{X}_{1}|U)}$ $\hat{X}_{1}^{n}$
sequences i.i.d. with pmf $p(\hat{x}_{1}|u)$ and a codebook of $2^{nI(X;\hat{X}_{2}|U)}$
sequences $\hat{X}_{2}^{n}$ i.i.d. with pmf $p(\hat{x}_{2}|u)$.
Given a source sequence $X^{n}$, the encoder finds a jointly typical
$U^{n}$ codeword, and then a codeword $\hat{X}_{1}^{n}$ jointly
typical with $(X^{n},U^{n})$ and similarly for $\hat{X}_{2}^{n}$.
Using source-channel separation on the broadcast {}``action'' channel
described above, the index from the $U$-codebook and a part of the
index from the $\hat{X}_{1}$-codebook, of rate $r$, is described
to both decoders, and the index from $\hat{X}_{2}$-codebook is described
to Decoder 2 as its private information. Thus, we have the inequalities
\begin{subequations} \label{rate} 
\begin{eqnarray}
\tilde{R}_{1} & \ge & I(X;U)+r\\
\textrm{and }\tilde{R}_{2} & \ge & I(X;\hat{X}_{2}|U).
\end{eqnarray}
 \end{subequations}The capacity region of the broadcast channel is
given by the conditions \cite[Chapter 9]{Elgammal}, $\tilde{R}_{1}\le H(\textrm{f}_{Y}(A))$
and $\tilde{R}_{1}+\tilde{R}_{2}\le H(A)$, and thus the following
rates are achievable \begin{subequations} \label{cap} 
\begin{eqnarray}
\tilde{R}_{1} & \le & H(\textrm{f}_{Y}(A))\\
\textrm{and }\tilde{R}_{2} & \le & H(A|\textrm{f}_{Y}(A)).
\end{eqnarray}
 \end{subequations} Finally the remaining part of the index of codeword
$\hat{X}_{1}^{n}$ is sent through the direct rate $R$, leading to
the condition 
\begin{eqnarray}
R\ge I(X;\hat{X}_{1}|U)-r.\label{r}
\end{eqnarray}
 Combining (\ref{rate}), (\ref{cap}) and (\ref{r}), and using Fourier-Motzkin
elimination we obtain 
\begin{eqnarray}
R & \ge & I(X;\hat{X}_{1},U)-H(\textrm{f}_{Y}(A))
\end{eqnarray}
 and (\ref{eq:enc_req11})-(\ref{eq:enc_req2}). The distortion and
cost constraints are handled in a standard manner and hence the details
are omitted.

It remains to discuss how to handle the case of causal or strictly
causal action observation. Given the converse result in Appendix C,
it is enough to show that (\ref{eq:R_NCC_enc})-(\ref{eqn: action_const_enc})
is achievable also with strictly causal and causal action observation.
This can be simply accomplished by encoding in blocks as per achievability
of Proposition \ref{prop:RD_SCC} and Proposition \ref{prop:RD_CC}.
Specifically, in each block the encoder compresses the source sequence
corresponding to the next block. Decoder 2 then operates as above,
while Decoder 1 can recover all source blocks at the end of all blocks.\hfill{}$\Box$ 
\begin{rem}
The scenario solved above is when the action observation is perfect,
i.e., $\textrm{f}(A)=A$. The result also carries verbatim for the
more general case where $\textrm{f}(A)$ is a generic function as
long as $H(\textrm{f}_{Y}(A)|\textrm{f}(A))=0$. The expressions of
the rate region remain the same as in the proposition above except
that $A$ is replaced by $\textrm{f}(A)$. 
\end{rem}
Proposition \ref{prop:RD_E_NCC} characterizes the optimal performance
for the case when Decoder 2 has a better information about the actions
taken by the encoder than Decoder 1 in the sense that $H(\textrm{f}_{Y}(A)\textrm{|f}(A))=0$.
We note here that a similar characterization can be given also for
the dual setting in which $H(\textrm{f}(A)\textrm{|f}_{Y}(A))=0$
so that Decoder 1 has the better observation about the actions. 
\begin{prop}
\label{prop_new}The rate-distortion-cost function $\mathcal{\mbox{\ensuremath{R(D_{1},D_{2},\Gamma)}}}$
for the source coding problem with encoder-side actions, non-causal,
causal or strictly causal observation of the actions illustrated in
Fig. \ref{fig:fig4} with $H(\ce{f}(A)|\ce{f}_{Y}(A))=0$, is given
by

\begin{eqnarray}
R(D_{1},D_{2},\Gamma)=\underset{p(a),\textrm{ }p(\hat{x}_{1},\hat{x}_{2}|x)}{\min}\{I(X;\hat{X}_{1},\hat{X}_{2})-H(\ce{f}_{Y}(A))\}^{+},\label{eq:R_NCC_enc_rev}
\end{eqnarray}
 where the information measures are evaluated with respect to the
joint pmf 
\begin{align}
p(x,\hat{x}_{1},\hat{x}_{2},a)=p(x)p(\hat{x}_{1},\hat{x}_{2}|x)p(a) & ,\label{eq:joint_enc_rev}
\end{align}
 such that the following inequalities are satisfied, \begin{subequations}\label{eqn: action_const_enc_rev}
\begin{eqnarray}
\ce{E}[d_{j}(X,\hat{X}_{j})] & \leq & D_{j},\textrm{ for \ensuremath{j=1,2,}}\\
\ce{E}[\Lambda(A)] & \leq & \Gamma,\\
I(X;\hat{X}_{2}) & \leq & H(\ce{f}(A)).\label{eq:enc_req11}
\end{eqnarray}
 \end{subequations} 
\end{prop}
The converse follows similarly as that for Proposition \ref{prop:RD_E_NCC}
where instead of $U$ in the converse we use $\hat{X}_{2}$, as knowing
$Y^{n}=\textrm{f}_{Y}(A^{n})$ implies knowing $\textrm{f}(A^{n})$,
due to the assumption $H(\textrm{f}(A)|\textrm{f}_{Y}(A))=0$. We
just outline the achievability for the non-causal case (the achievability
for strictly causal and causal case uses block coding ideas as in
Proposition \ref{prop:RD_E_NCC}). A successive refinement codebook
is generated by drawing $2^{nI(X;\hat{X}_{2})}$ codewords $\hat{X}_{2}^{n}$,
and, for each codeword $\hat{X}_{2}^{n}$, a number $2^{nI(X;\hat{X}_{1}|\hat{X}_{2})}$
of codewords $\hat{X}_{1}^{n}$. As for Proposition \ref{prop:RD_E_NCC},
the indices of these two codebooks obtained via standard joint typicality
encoding are sent through the degraded broadcast channel $p(y,b|a)=\mathbf{1}_{\{y=\ce{f}_{Y}(A),b=\ce{f}(A)\}}$.
Splitting the rate for the index of codeword $\hat{X}_{1}^{n}$ so
that a rate $R$ is sent over the direct link to Decoder 1, reliability
of compression and communication over the {}``action'' broadcast
channel is guaranteed if 
\begin{eqnarray}
I(X;\hat{X}_{2}) & \le & H(\ce{f}(A))\\
I(X;\hat{X}_{2})+I(X;\hat{X}_{1}|\hat{X}_{2})-R & \le & H(\ce{f}(A),\ce{f}_{Y}(A))=H(\ce{f}_{Y}(A)),
\end{eqnarray}
 where the latter inequality implies $R\ge I(X;\hat{X}_{1},\hat{X}_{2})-H(\ce{f}_{Y}(A))$.
The proof is concluded using the usual steps.\hfill{}$\Box$

\section{Actions for Channel State Control and Probing\label{sec:Actions-for-Channel}}

In this section, we consider the impact of information embedding on
actions for the set-up of channel coding with actions of \cite{Weissman}.
To this end, we consider the model in Fig. \ref{fig:fig6}, in which
Decoder 1, based on the observation of a deterministic function of
the actions, wishes to retrieve part of the information destined to
Decoder 2. Note that for simplicity of notation here the additional
encoder that observes the actions is denoted as Decoder 1, rather
than Decoder 2 as done above. Also, we emphasize that in the original
set-up of \cite{Weissman}, Decoder 1 was not present.

\subsection{System Model\label{sub:System-model_act}}

The system is defined by the pmfs $p(x)$, $p(y|x,s,a)$, $p(s|a)$,
function $\textrm{f}\textrm{: }\mathcal{A}\rightarrow\mathcal{B}$
and by discrete alphabets $\mathcal{X},\mathcal{A\mbox{, \ensuremath{\mathcal{B}}, }S}$,
and ${\cal Y}$. Given the messages $(M_{1},M_{2})$, selected randomly
from the set $\mathcal{M}_{1}\times\mathcal{M}_{2}=[1,2^{nR_{1}}]\times[1,2^{nR_{2}}]$,
an action sequence $A^{n}\in\mathcal{A}^{n}$ is selected by the Encoder.
Decoder 1 observes the signal $B^{n}=\textrm{f}(A^{n})$ as a deterministic
function of the actions, and estimates message $M_{1}$. Note that
the notation here implies a {}``non-causal'' observation of the
actions, but it is easy to see that the results below hold also with
causal and strictly causal observation of the actions. Moreover, the
state sequence $S^{n}\in\mathcal{S}^{n}$ is generated as the output
of a memoryless channel $p(s|a)$ and we have $p(b^{n},s^{n}|a^{n})=\prod_{i=1}^{n}p(s_{i}|a_{i})\mathbf{1}_{\{b_{i}=\textrm{f}(a_{i})\}}$
for an action sequence $A^{n}=a^{n}$. The input sequence $X^{n}\in\mathcal{X}^{n}$
is selected on the basis of both messages $(M_{1},M_{2})$ and of
the state sequence $S^{n}$ by the Encoder. The action sequence $A^{n}$
and the input $X^{n}$ have to satisfy an average cost constraint
defined by a function $\gamma:\mathcal{A}\times\mathcal{X}\rightarrow[0,\infty)$,
so that the cost for the input sequences $a^{n}$ and $x^{n}$ is
given by $\gamma(a^{n},x^{n})=\frac{1}{n}\sum_{i=1}^{n}\gamma(a_{i},x_{i}).$
Given $X^{n}=x^{n}$, $S^{n}=s^{n}$ and $A^{n}=a^{n}$, the received
signal is distributed as $p(y^{n}|x^{n},s^{n},a^{n})=\prod_{i=1}^{n}p(y_{i}|x_{i},s_{i},a_{i})$.
Decoder 2, having received the signal $Y^{n}$, estimates both messages
$(M_{1},M_{2})$.

The setting includes the semi-deterministic broadcast channel with
degraded message sets \cite{KornerMarton} (see also \cite[Ch. 8]{Elgammal})
as a special case by setting $X$ to be constant and $Y=S$, and the
channel with action-dependent states studied in \cite{Weissman} for
$R_{1}=0$. 
\begin{defn}
\label{def_pp}An $(n,R_{0},R_{1},\Gamma,\epsilon)$ code for the
model in Fig. \ref{fig:fig6} consists of an action encoder 
\begin{align}
\textrm{h}^{(a)}\textrm{: }\mathcal{M}_{1}\times\mathcal{M}_{2}\rightarrow\mathcal{A}^{n},\label{eq:action_enc}
\end{align}
 which maps message $(M_{1},M_{2})$ into an action sequence $A^{n}$;
a channel encoder 
\begin{align}
\textrm{h}^{(e)}\textrm{: }\mathcal{M}_{1}\times\mathcal{M}_{2}\times\mathcal{S}^{n}\rightarrow\mathcal{X}^{n},\label{eq:enc}
\end{align}
 which maps message $\mbox{\ensuremath{(M_{1},M_{2})}}$ and the state
sequence $S^{n}$ into the sequence $X^{n}$; two decoding functions
\begin{align}
\textrm{h}_{1}^{(d)}\textrm{: }\mathcal{B}^{n}\rightarrow\mathcal{M}_{1},\\
\textrm{and }\textrm{h}_{2}^{(d)}\textrm{: }\mathcal{Y}^{n}\rightarrow\mathcal{M}_{1}\times\mathcal{M}_{2},
\end{align}
 which map the sequences $B^{n}$ and $Y^{n}$ into the estimated
messages $\hat{M}_{1}$ and $(\hat{M}_{1},\hat{M}_{2})$, respectively;
such that the probability of error in decoding the messages $(M_{1},M_{2})$
is small, 
\begin{align}
\textrm{Pr}[\textrm{h}_{1}^{(d)}(B^{n})\neq M_{1}] & \leq\epsilon,\\
\textrm{and }\textrm{Pr}[\textrm{h}_{2}^{(d)}(Y^{n})\neq(M_{1},M_{2})] & \leq\epsilon,
\end{align}
 and the cost constraint is satisfied, i.e., 
\begin{eqnarray}
\frac{1}{n}\sum_{i=1}^{n}\textrm{E}\left[\gamma(A_{i},X_{i})\right] & \leq & \Gamma+\epsilon.\label{eq:cost_const}
\end{eqnarray}

Given a cost $\Gamma$, a rate pair $(R_{1},R_{2})$ is said to be
achievable for a cost-constraint $\Gamma$ if, for any $\epsilon>0$
and sufficiently large $n$, there a exists a $(n,R_{1},R_{2},\Gamma,\epsilon)$
code. We are interested in characterizing the capacity-cost region
${\cal C}(\Gamma)$, which is the closure of all achievable rate pairs
$(R_{1},R_{2})$ for the given cost $\Gamma$. 
\end{defn}

\subsection{Capacity-Cost Region\label{sub:Capacity_Fun}}

In this section, a single-letter characterization of the capacity-cost
region is derived. 
\begin{prop}
\label{prop:act_stat}The capacity-cost region ${\cal C}(\Gamma)$
for the system in Fig. \ref{fig:fig6} is given by the union of all
rate pairs $(R_{1},R_{2})$ such that the inequalities \begin{subequations}\label{eqn: channel_region}
\begin{eqnarray}
R_{1} & \leq & H(\ce{f}(A))\label{eq:R1}\\
\ce{and}\textrm{ }R_{1}+R_{2} & \leq & I(A,U;Y)-I(U;S|A),\label{eq:R1+R2}
\end{eqnarray}
 \end{subequations} are satisfied, where the mutual informations
are evaluated with respect to the joint pmf 
\begin{align}
p(a,s,u,x,y)=p(a)p(s|a)p(u|s,a)\mathbf{1}_{\{x=\ce{g}(u,s)\}}p(y|x,s,a) & ,\label{eq:joint_channel}
\end{align}
 for some pmfs $p(a)$, $p(u|s,a)$ and function $\ce{g}\textrm{: }\mathcal{U}\times\mathcal{S}\rightarrow{\cal X}$
such that\textup{ 
\begin{eqnarray}
\mathrm{E}[\gamma(A,X)] & \leq & \Gamma.\label{eq:cost_bound}
\end{eqnarray}
 }Finally, we can set $|\mathcal{U}|\leq|\mathcal{X}||\mathcal{S}||\mathcal{A}|+1$
without loss of optimality. 
\end{prop}
\begin{figure}[h!]
\centering\includegraphics[bb=37bp 485bp 571bp 752bp,clip,scale=0.65]{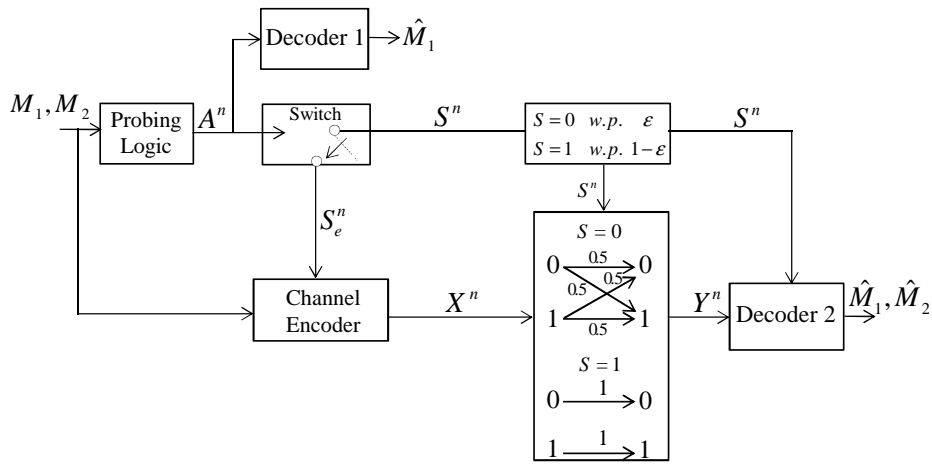}
\caption{Channel coding with actions for channel state probing and with information
embedding on actions.}

\label{fig:channel_ex} 
\end{figure}

The proof of converse is an immediate consequence of cut-set arguments
and of the proof of the upper bound obtained in \cite[Theorem 1]{Weissman}.
Specifically, inequality (\ref{eq:R1}) follows by considering the
cut around Decoder 1, while the inequality (\ref{eq:R1+R2}) coincides
with the bound derived in \cite[Theorem 1]{Weissman} on the rate
that can be communicated between the Encoder and Decoder 2 with no
regards for Decoder 1%
\footnote{The cardinality constraints follow from \cite[Theorem 1]{Weissman}%
}. The achievability requires rate splitting, superposition coding
and the coding strategy proposed in \cite[Theorem 1]{Weissman}. A
sketch of proof of the achievability is relegated to Appendix D.

\subsection{Probing Capacity\label{sub:Probing-Capacity}}

Here we provide an example that illustrates the effect of the communication
requirements of the action-cribbing decoder on the system performance.
Consider the communication system shown in Fig. \ref{fig:channel_ex},
where the states is known to Decoder 2. We further assume binary actions,
such that, if $A=1$, the channel encoder observes the state $S$,
and if $A=0$, it does not obtain any information about $S$. We model
this problem by defining the state information available at the encoder
as $S_{e}=\textrm{u}(S,A)$, where $\textrm{u}(S,1)=S$ and $\textrm{u}(S,0)=\textrm{e}$,
where represents as {}``erasure'' symbol. Following \cite{probing},
we refer to this problem as having a {}``probing'' encoder.

The channel encoder maps the state information $S_{e}^{n}$ and messages
$M_{1},M_{2}$ into a codeword $X^{n}$ (see Fig. \ref{fig:channel_ex}).
Moreover, two cost constraints, namely $\frac{1}{n}\sum_{i=1}^{n}\textrm{E}\left[\gamma_{a}(A_{i})\right]\leq\Gamma_{A}$
and $\frac{1}{n}\sum_{i=1}^{n}\textrm{E}\left[\gamma_{x}(X_{i})\right]\leq\Gamma_{X}$
are imposed for given action input cost functions $\gamma_{a}:\mathcal{A}\rightarrow[0,\Lambda_{a,max}]$
and $\gamma_{x}:\mathcal{X}\rightarrow[0,\Lambda_{x,max}]$ with $0\leq\Lambda_{a,max}<\infty$
and $0\leq\Lambda_{x,max}<\infty$, respectively. In \cite[Theorem 1]{probing},
a correspondence was proved between the set-up of a probing encoder
and that of action dependent states. Using \cite[Theorem 1]{probing}
and Proposition \ref{prop:act_stat}, we can easily obtain that the
capacity-cost region ${\cal C}(\Gamma_{A},\Gamma_{X})$ for the system
in Fig. \ref{fig:channel_ex} is given by the union of all rate pairs
$(R_{1},R_{2})$ such that the inequalities\begin{subequations}\label{eqn: channel_region_ex}
\begin{eqnarray}
R_{1} & \leq & H(A|Q)\label{eq:R1_ex}\\
\ce{and}\textrm{ }R_{1}+R_{2} & \leq & I(X;Y|S,Q),\label{eq:R1+R2_ex}
\end{eqnarray}
 \end{subequations} are satisfied, where the mutual informations
are evaluated with respect to the joint pmf 
\begin{align}
p(q,a,s,s_{e},x,y)=p(q)p(a|q)p(s)\mathbf{1}_{\{s_{e}=\ce{u}(s,a)\}}p(x|s_{e},a,q)p(y|x,s) & ,\label{eq:joint_channel-1}
\end{align}
 for some pmfs $p(q)$, $p(a|q)$, $p(x|s_{e},a,q)$ such that $\mathrm{E}[\gamma_{a}(A)]\leq\Gamma_{A}$
and $\mathrm{E}[\gamma_{x}(X)]\leq\Gamma_{X}$.

We now apply (\ref{eq:R1_ex})-(\ref{eq:R1+R2_ex}) to the channel
shown in Fig. \ref{fig:channel_ex} in which alphabets are binary
${\cal X}={\cal Y}={\cal S}=\{0,1\}$, $S$ is a $\textrm{Bern}(1-\epsilon)$
variable for $0\leq\epsilon\leq1$ and the channel is a binary symmetric
with flipping probability $0.5$ if $S=0$ ({}``bad'' channel state)
and $0$ if $S=1$ ({}``good'' channel state).

To evaluate the maximum achievable sum-rate $R_{1}+R_{2}$ for a given
rate $R_{1}$, we define $\ce{Pr}[A=1]=\gamma$, $\ce{Pr}[X=1|S_{e}=1,A=1]=p_{1}$
and $\ce{Pr}[X=1|S_{e}=\ce{e},A=0]=p_{2}$, and we set $\ce{Pr}[X=1|S_{e}=0,A=1]=0$
without loss of optimality. The maximum sum-rate $R_{1}+R_{2}$ for
a given rate $R_{1}$ is then obtained from (\ref{eq:R1+R2_ex}) by
solving the problem 
\begin{eqnarray}
R_{1}+R_{2} & = & \underset{0\leq p_{1},p_{2},\gamma\leq1}{\ce{max}}\gamma(1-\epsilon)H(p_{1})+(1-\gamma)(1-\epsilon)H(p_{2}),\label{eq:opt_ex}
\end{eqnarray}
 under the constraint $\textrm{E}[X]=p_{1}\gamma(1-\epsilon)+p_{2}(1-\gamma)\leq\Gamma_{X}$,
$\textrm{E}[A]=\gamma\leq\Gamma_{A}$ and $H(A)=H(\gamma)\geq R_{1}$.
Note that the last constraint imposes that the rate achievable by
the Decoder 1 is larger than $R_{1}$ as per (\ref{eq:R1_ex}).

\begin{figure}[h!]
\centering\includegraphics[bb=37bp 345bp 542bp 735bp,clip,scale=0.6]{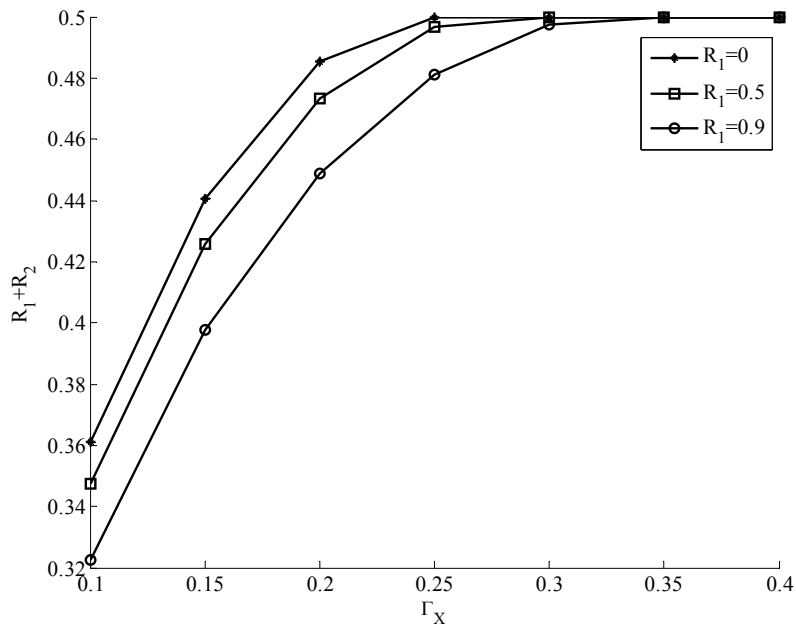}
\caption{Sum-rate $R_{1}+R_{2}$ versus the input cost constraint $\Gamma_{X}$
for values of $R_{1}=0$, $R_{1}=0.5$ and $R_{1}=0.9$.}

\label{fig:plot_channel_prob} 
\end{figure}

The sum-rate in (\ref{eq:opt_ex}) is shown in Fig. \ref{fig:plot_channel_prob}
for $\epsilon=0.5$, $\Gamma_{A}=1$ and different values of $R_{1}$.
It can be seen that, for sufficiently small values of the cost constraint
$\Gamma_{X},$ increasing the communication requirements, i.e., $R_{1}$,
of the Decoder 1, reduces the achievable sum-rate $R_{1}+R_{2}$.
This is due to the fact that increasing $R_{1}$ requires to encode
more information in the action sequence, which in turn reduces the
portion of the actions that can be set to $A=1$, i.e., $\textrm{Pr}[A=1]$.
As a result, the encoder is less informed about the state sequence
and thus bound to waste some power on bad channel states.

\begin{rem}
The communication requirements of Decoder 1 need not necessarily affect
the system performance. For instance, consider the example 1 in \cite[Sec. V.A]{probing},
which includes a probing encoder as in Fig. \ref{fig:channel_ex}
but transmitting over a different channel. There, it turns out that
it is sufficient to have $\textrm{Pr}[A=1]\gtrsim0.2$ in order to
achieve the same performance that can be achieved with full encoder
channel state information. Therefore, the additional constraint on
the rate of Decoder 1 (\ref{eq:R1_ex}), namely $R_{1}\geq H(A)$,
does not affect the sum-rate achievable in this example for any rate
$R_{1}\in[0,1]$. 
\end{rem}

\section{Concluding Remarks}

There is a profound interplay between actuation and communication
in that both actuation can be instrumental to improve the efficiency
of communication, and, vice versa, communication, implicit or explicit,
can provide an essential tool to improve control tasks. This work
has focused on the first type of interplay, and has investigated the
implications of embedding information directly in the actions for
the aim of communicating with a separate decoder. The communication
requirements of this decoder are generally in conflict with the goal
of improving the efficiency of the given communication link. This
performance trade-off has been studied here for both source and channel
coding. The results provided in the paper allow to give a quantitative
answer to the questions posed in Sec. \ref{sub:Information-Embedding-on}
regarding the impact of the requirements of action information embedding
on the system performance. They also shed light into the structure
of optimal embedding strategies, which turns out to be related, for
the source coding model, with the strategies studied in \cite{Weissman_crib,cuff zhao}.

The investigation on the theme of information embedding on actions
can be further developed in a number of directions, including models
with memory \cite{asnani feedback,Asnani real time} and with multiple
terminals \cite{Weissman_multi,chiru behzad conf,Zhao compression with action}.
We also note that results akin to the ones reported here can be developed
assuming causal state information at the decoder for source coding
problems or causal state information at the transmitter.

\appendices{ }

\section*{Appendix A: Proof of Proposition \ref{prop:RD_NCC}}

Here, we prove the converse part of Proposition \ref{prop:RD_NCC}.
For any $(n,R,D_{1}+\epsilon,D_{2}+\epsilon,\Gamma+\epsilon)$ code,
we have 
\begin{eqnarray}
nR & \geq & H(M)\nonumber \\
 & = & I(M;X^{n},Y^{n})\nonumber \\
 & = & H(X^{n},Y^{n})-H(X^{n},Y^{n}|M)\nonumber \\
 & = & H(X^{n})+H(Y^{n}|X^{n})-H(Y^{n}|M)-H(X^{n}|M,Y^{n})\nonumber \\
 & = & \negmedspace\negmedspace\overset{n}{\underset{i=1}{\sum}}H(X_{i})+H(Y_{i}|Y^{i-1},X^{n})-H(Y_{i}|Y^{i-1},M)-H(X_{i}|X^{i-1},M,Y^{n})\nonumber \\
 & \overset{(a)}{\geq} & \overset{n}{\negmedspace\negmedspace\underset{i=1}{\sum}}H(X_{i})\negmedspace+\negmedspace H(Y_{i}|Y^{i-1},X^{n},A^{n})\negmedspace-\negmedspace H(Y_{i}|Y^{i-1},M,A^{n})\negmedspace-\negmedspace H(X_{i}|X^{i-1},M,Y^{n},A^{n})\label{eq:insertion of A}\\
 & \overset{(b)}{=} & \overset{n}{\negmedspace\negmedspace\underset{i=1}{\sum}}H(X_{i})\negmedspace+\negmedspace H(Y_{i}|Y^{i-1},X^{n},A^{n},\hat{X}_{2i})\negmedspace-\negmedspace H(Y_{i}|Y^{i-1},M,A^{n},\hat{X}_{2i})\negmedspace\label{eq:insertion of X^2}
\end{eqnarray}

\begin{eqnarray}
 &  & -H(X_{i}|X^{i-1},M,Y^{n},A^{n},\hat{X}_{2i})\nonumber \\
 & \overset{(c)}{\geq} & \negmedspace\negmedspace\overset{n}{\underset{i=1}{\sum}}H(X_{i})+\negmedspace H(Y_{i}|X_{i},A_{i},\hat{X}_{2i})-\negmedspace H(Y_{i}|A_{i},\hat{X}_{2i})-H(X_{i}|U_{i},Y_{i},A_{i},\hat{X}_{2i}),\label{eq:conv_end1}
\end{eqnarray}
where (\textit{a}) because $A^{n}$ is a function of $M$ and since
conditioning reduces entropy; ($b$) follows since $\hat{X}_{2i}$
is a function of $A^{n}$; and ($c$) follows because we have the
Markov relation $Y_{i}\textrm{---}(X_{i},A_{i},\hat{X}_{2i})\textrm{---}(X^{n\backslash i},A^{n\backslash i})$,
by defining $U_{i}=(M,X^{i-1},Y^{n\backslash i},A^{i-1})$ and since
conditioning decreases entropy.

Defining $Q$ to be a random variable uniformly distributed over $[1,n]$
and independent of all the other random variables and with $X\overset{\triangle}{=}X_{Q}$,
$Y\overset{\triangle}{=}Y_{Q}$, $A\overset{\triangle}{=}A_{Q}$,
$\hat{X}_{1}\overset{\triangle}{=}\hat{X}_{1Q}$, $\hat{X}_{2}\overset{\triangle}{=}\hat{X}_{2Q}$
and $U\overset{\triangle}{=}(U_{Q},Q),$ from (\ref{eq:conv_end1})
we have 
\begin{eqnarray*}
nR & \geq & H(X|Q)+\negmedspace H(Y|X,A,\hat{X}_{2},Q)\negmedspace-H(Y|A,\hat{X}_{2},Q)-H(X|U,Y,A,\hat{X}_{2},Q)\\
 & \overset{(a)}{\geq} & H(X)+\negmedspace H(Y|X,A,\hat{X}_{2})\negmedspace-H(Y|A,\hat{X}_{2})-H(X|U,Y,A,\hat{X}_{2})\\
 & = & I(X;U,Y,A,\hat{X}_{2})-I(Y;X|A,\hat{X}_{2})\\
 & = & I(X;A,\hat{X}_{2})+I(X;U|Y,A,\hat{X}_{2}),
\end{eqnarray*}
 where in ($a$) we have used the fact that $X^{n}$ is i.i.d., conditioning
reduces entropy and by the problem definition. Moreover, we have the
following chain of inequalities 
\begin{eqnarray}
H(\textrm{f}(A^{n})) & \leq & \overset{n}{\underset{i=1}{\sum}}H(\textrm{f}(A_{i}))=nH(\textrm{f}(A)|Q)\leq nH(\textrm{f}(A)),\label{eq:reqU}
\end{eqnarray}
 where the last inequality follows since conditioning reduces entropy,
and
\begin{eqnarray*}
H(\textrm{f}(A^{n})) & \geq & I(\textrm{f}(A^{n});X^{n})\\
 & = & \overset{n}{\underset{i=1}{\sum}}I(\textrm{f}(A^{n});X_{i}|X^{i-1})\\
 & = & \overset{n}{\underset{i=1}{\sum}}I(\textrm{f}(A^{n}),\hat{X}_{2i};X_{i}|X^{i-1})\\
 & = & \overset{n}{\underset{i=1}{\sum}}I(\textrm{f}(A^{n}),\hat{X}_{2i},X^{i-1};X_{i})
\end{eqnarray*}
\begin{eqnarray}
 & \overset{(a)}{\geq} & \overset{n}{\underset{i=1}{\sum}}I(\textrm{f}(A_{i}),\hat{X}_{2i};X_{i})\nonumber \\
 & \overset{}{=} & n(H(X|Q)-H(X|\textrm{f}(A),\hat{X}_{2},Q))\nonumber \\
 & \overset{(b)}{\geq} & n(H(X)-H(X|\textrm{f}(A),\hat{X}_{2}))\nonumber \\
 & = & n(I(X;\textrm{f}(A),\hat{X}_{2})),\label{eq:reqL}
\end{eqnarray}
 where ($a$) follows by the chain for mutual information and since
mutual information is non-negative; and ($b$) follows since $X^{n}$
is i.i.d. and due to the fact that conditioning decreases entropy.
Combining (\ref{eq:reqL}) and (\ref{eq:reqU}), we obtain the inequality
\begin{eqnarray}
I(X;\textrm{f}(A),\hat{X}_{2}) & \leq & H(\textrm{f}(A)).\label{eq:req}
\end{eqnarray}

We note that the defined random variables factorizes as (\ref{eq:joint})
since we have the Markov chain relationship $(\hat{X}_{1},\hat{X}_{2},U)$---$(A,X)$---$Y$
by the problem definition and that $\hat{X}_{2}$ is a function $\textrm{g}(U,Y)$
of $U$ and $Y$ by the definition of $U$. Moreover, from cost and
distortion constraints (\ref{action cost})-(\ref{dist const}), we
have\begin{subequations}\label{eqn: action_cascade_const-1} 
\begin{align}
D_{j}+\epsilon & \geq\textrm{\ensuremath{\frac{1}{n}\sum_{i=1}^{n}}E}[d_{j}(X_{i},\hat{X}_{ji})]=\textrm{E}[d_{j}(X,\hat{X}_{j})],\textrm{ for }j=1,2,\label{eq:dist1-1}\\
\textrm{and }\Gamma+\epsilon & \geq\frac{1}{n}\sum_{i=1}^{n}\textrm{E}\left[\Lambda(A_{i})\right]=\textrm{E}\left[\Lambda(A)\right].
\end{align}
 \end{subequations}

The cardinality constraint on the auxiliary random variable $U$ is
obtained as follows using Caratheodory's theorem as in \cite[Appendix C]{Elgammal}.
Note that we can write $I(X;\hat{X}_{2},A)+I(X;U|\hat{X}_{2},A,Y)=H(X)-H(X|\hat{X}_{2},A)+H(X|\hat{X}_{2},A,Y)-H(X|\hat{X}_{2},A,Y,U)$.
Now, to preserve the joint distribution of variables ($X$,$\hat{X}_{2}$,$A$),
and thus the distribution of all variables ($X$,$\hat{X}_{2}$,$A$,$Y$)
and the terms $H(X),\textrm{ }H(X|\hat{X}_{2},A)$ and $H(X|\hat{X}_{2},A,Y)$
(since $p(y|x,a)$ is fixed), the set $\mathcal{U}$ should have $|\mathcal{X}||\hat{\mathcal{X}}_{2}||\mathcal{A}|-1$
elements; moreover, we need one further element to preserve the conditional
entropy $H(X|\hat{X}_{2},A,Y,U)$ and one for the distortion $E[d_{1}(X,\hat{X}_{1})]$.

\section*{Appendix B: Proof of Proposition \ref{prop:RD_SCC} and Proposition
\ref{prop:RD_CC}}

Here, we first prove the converse part of Proposition \ref{prop:RD_SCC}
and then describe the different steps needed to prove Proposition
\ref{prop:RD_CC}. The first part of the converse follows the same
steps as in Appendix A. However, we note that in (\ref{eq:insertion of A})
and (\ref{eq:insertion of X^2}), we can write $A^{i}$ instead of
$A^{n}$, without changing the following steps. This is due to the
strictly causal dependence of $\hat{X}_{2i}$ on the action sequence
which is used in (\ref{eq:insertion of X^2}). This allows to validate
the claim in Remark \ref{adaptive_action}. To prove the constraint
in (\ref{eq:req_SCC}), we have the following chain of inequalities
\begin{eqnarray}
H(\textrm{f}(A^{n})) & = & \overset{n}{\underset{i=1}{\sum}}H(\textrm{f}(A_{i})|\textrm{f}(A^{i-1}))=\overset{n}{\underset{i=1}{\sum}}H(\textrm{f}(A_{i})|\hat{X}_{2i})\overset{(a)}{\leq}nH(\textrm{f}(A)|\hat{X}_{2}).\label{eq:reqU-1}
\end{eqnarray}
 Moreover, we can write 
\begin{eqnarray}
H(\textrm{f}(A^{n})) & \geq & I(\textrm{f}(A^{n});X^{n})\nonumber \\
 & = & \overset{n}{\underset{i=1}{\sum}}I(\textrm{f}(A^{n});X_{i}|X^{i-1})\nonumber \\
 & = & \overset{n}{\underset{i=1}{\sum}}I(\textrm{f}(A^{n}),\hat{X}_{2i};X_{i}|X^{i-1})\nonumber \\
 & = & \overset{n}{\underset{i=1}{\sum}}I(\textrm{f}(A^{n}),\hat{X}_{2i},X^{i-1};X_{i})\nonumber \\
 & \overset{(a)}{\geq} & \overset{n}{\underset{i=1}{\sum}}I(\textrm{f}(A_{i}),\hat{X}_{2i};X_{i})\nonumber \\
 & \overset{}{=} & n(H(X|Q)-H(X|\textrm{f}(A),\hat{X}_{2},Q))\nonumber \\
 & \overset{(b)}{\geq} & n(H(X)-H(X|\textrm{f}(A),\hat{X}_{2}))\nonumber \\
 & = & n(I(X;\textrm{f}(A),\hat{X}_{2})),\label{eq:reqL-1}
\end{eqnarray}
where ($a$) follows by the chain for mutual information and since
mutual information is non-negative; and ($b$) follows since $X^{n}$
is i.i.d. and due to the fact that conditioning decreases entropy.
Combining (\ref{eq:reqL-1}) and (\ref{eq:reqU-1}), we obtain the
inequality (\ref{eq:req_SCC}). We note that the joint pmf of the
defined random variables factorizes as (\ref{eq:joint_SCC}) since
we have the Markov chain relationship $(\hat{X}_{1},\hat{X}_{2},U)$---$(A,X)$---$Y$
by the problem definition and that $\hat{X}_{1}$ is a function $\textrm{g}(U,Y)$
of $U$ and $Y$ by the definition of $U$ as in Appendix A. The distortion,
cost and cardinality constraint are obtained as in Appendix A.

The converse for Proposition \ref{prop:RD_CC} follows from similar
steps by defining $V_{i}=\textrm{f}(A^{i-1})$ and noting that $\hat{X}_{2i}$
is a function of $V_{i}$ and $\textrm{f}(A_{i})$.

We bound the cardinality of the auxiliary random variables $U$ and
$V$ for Proposition \ref{prop:RD_CC} using \cite[Appendix C]{Elgammal}.
The bounds for $U$ in Proposition \ref{prop:RD_SCC} follow in the
same way. Note that we can write $I(X;V,A)+I(X;U|V,A,Y)=H(X)-H(X|V,A)+H(X|V,A,Y)-H(X|V,A,Y,U)$.
Starting with $V$, the alphabet $\mathcal{V}$ should have $|\mathcal{X}|-1$
elements to preserve the distribution $p(x)$ and hence $H(X)$, one
element to preserve $-H(X|V,A)+H(X|V,A,Y)$, two elements to preserve
the distortion constraints and and one more to preserve the condition
$I(X;V,\textrm{f}(A))\le H(\textrm{f}(A)|V)$. As for $U$, just as
in Appendix A, $U$ should have $|\mathcal{X}||\mathcal{V}||\mathcal{A}|-1$
elements to preserve the joint distribution $p(x,v,a)$ (which preserves
the joint distribution $p(x,a,v,y)$ and hence $H(X),\textrm{ }H(X|V,A),\textrm{ }H(X|V,A,Y)$),
one element to preserve $H(X|V,A,Y,U)$ and one more to preserve the
distortion constraint of Decoder 1.

\section*{Appendix C: Proof of Proposition \ref{prop:RD_E_NCC}}

Here, we prove the converse part of Proposition \ref{prop:RD_E_NCC}.
To establish the converse, it is sufficient to consider the case of
non-causal action observation, as done in the following. For any $(n,R,D_{1}+\epsilon,D_{2}+\epsilon,\Gamma+\epsilon)$
code, define the auxiliary variable $U_{i}=(Y^{n},X^{i-1})$, and
$Q$ as a random time sharing variable uniformly distributed in the
interval $[1,n]$ independent of $(X,U,\hat{X}_{1},\hat{X}_{2},A,Y)$.
We then have, 
\begin{eqnarray}
H(Y^{n}) & = & \sum_{i=1}^{n}H(Y_{i}|Y^{i-1})\nonumber \\
 & \le & \sum_{i=1}^{n}H(Y_{i})\nonumber \\
 & = & \sum_{i=1}^{n}H(\textrm{f}_{Y}(A_{i}))\nonumber \\
 & = & nH(\textrm{f}_{Y}(A_{Q})|Q)\nonumber \\
 & \le & H(\textrm{f}_{Y}(A_{Q})).\label{eq:eqy1}
\end{eqnarray}
 Also, we can write 
\begin{eqnarray}
H(Y^{n}) & \ge & I(X^{n};Y^{n})\nonumber \\
 & = & \sum_{i=1}^{n}I(X_{i};Y^{n}|X^{i-1})\nonumber \\
 & \stackrel{(a)}{=} & \sum_{i=1}^{n}I(X_{i};Y^{n},X^{i-1})\nonumber \\
 & = & \sum_{i=1}^{n}I(X_{i};U_{i})\nonumber \\
 & = & nI(X_{Q};U_{Q}|Q)\nonumber \\
 & \stackrel{(b)}{=} & I(X_{Q};U_{Q},Q),\label{eq:eqy2}
\end{eqnarray}
along with 
\begin{eqnarray}
H(A^{n}|Y^{n}) & = & \sum_{i=1}^{n}H(A_{i}|Y^{n},A^{i-1})\nonumber \\
 & \le & \sum_{i=1}^{n}H(A_{i}|Y_{i})\nonumber \\
 & = & \sum_{i=1}^{n}H(A_{i}|\textrm{f}_{Y}(A_{i}))\nonumber \\
 & = & n(H(A_{Q})|\textrm{f}_{Y}(A_{Q}),Q)\nonumber \\
 & \le & H(A_{Q}|\textrm{f}_{Y}(A_{Q})),\label{eq:eqa1}
\end{eqnarray}
and 
\begin{eqnarray}
H(A^{n}|Y^{n}) & \ge & I(X^{n};A^{n}|Y^{n})\nonumber \\
 & \stackrel{(c)}{=} & I(X^{n};A^{n},\hat{X}_{2}^{n}|Y^{n})\nonumber \\
 & \ge & \sum_{i=1}^{n}I(X_{i};\hat{X}_{2,i}|Y^{n},X^{i-1})\nonumber \\
 & = & \sum_{i=1}^{n}I(X_{i};\hat{X}_{2,i}|U_{i})\nonumber \\
 & = & nI(X_{Q};\hat{X}_{2,Q}|U_{Q},Q).\label{eq:eqa2}
\end{eqnarray}
Furthermore, we have 
\begin{eqnarray}
H(Y^{n},M) & \le & \sum_{i=1}^{n}H(Y_{i})+nR\nonumber \\
 & \le & nH(Y_{Q})+nR,\label{eq:eqr1}
\end{eqnarray}
 and
\begin{eqnarray*}
H(Y^{n},M) & \ge & I(X^{n};Y^{n},M)\\
 & \stackrel{(d)}{=} & I(X^{n};\hat{X}_{1}^{n},Y^{n},M)\\
 & \ge & \sum_{i=1}^{n}I(X_{i};\hat{X}_{1,i},Y^{n}|X^{i-1})\\
 & = & \sum_{i=1}^{n}I(X_{i};\hat{X}_{1,i},Y^{n},X^{i-1})
\end{eqnarray*}
\begin{eqnarray}
 & = & \sum_{i=1}^{n}I(X_{i};\hat{X}_{1,i},U_{i})\nonumber \\
 & = & n(X_{Q};\hat{X}_{1,Q},U_{Q},Q),\label{eq:eqr2}
\end{eqnarray}
where ($a$) follows from the independence of $X_{i}$ and $X^{i-1}$;
($b$) follows from the independence of $Q$ from all other random
variables; ($c$) follows from the fact that $\hat{X}_{2}^{n}$ is
a function of $A^{n}$; and ($d$) follows from the fact that $\hat{X}_{1}^{n}$
is a function of $(M,Y^{n})$. Defining $U\overset{\triangle}{=}(U_{Q},Q)$
along with $X\overset{\triangle}{=}X_{Q}$, $Y\overset{\triangle}{=}Y_{Q}$,
$A\overset{\triangle}{=}A_{Q}$, $\hat{X}_{1}\overset{\triangle}{=}\hat{X}_{1Q}$,
$\hat{X}_{2}\overset{\triangle}{=}\hat{X}_{2Q}$ and combining (\ref{eq:eqy1}),
(\ref{eq:eqy2}), (\ref{eq:eqa1}), (\ref{eq:eqa2}), (\ref{eq:eqr1})
and (\ref{eq:eqr2}), we obtain the rate region inequalities as mentioned
in the proposition. Note that the joint distribution of the random
variables $(X,Y,A,\hat{X}_{1},\hat{X}_{2})$ established above factorizes
as $p(x)p(u,\hat{x}_{1},\hat{x}_{2},a|x)$ but can be restricted only
to pmfs factorizing as in (\ref{eq:joint_enc}). This is because the
information measures in (\ref{eq:R_NCC_enc})-(\ref{eqn: action_const_enc})
only depends on the marginals $p(x,u,\hat{x}_{1})$, $p(a)$ and $p(x,u,\hat{x}_{2})$.
Distortion and cost constraints are handled in the standard manner
\cite{Elgammal}.

We bound the cardinality of the auxiliary random variables $U$ using
\cite[Appendix C]{Elgammal}. The set $\mathcal{U}$ should have $|\mathcal{X}||\mathcal{\hat{X}}_{1}||\mathcal{\hat{X}}_{2}|-1$
elements to preserve the joint distribution $p(x,\hat{x}_{1},\hat{x}_{2})$,
one element to preserve the Markov chain $\hat{X}_{1}-U-\hat{X}_{2}$,
and three elements to preserve $H(X|\hat{X}_{1},U)$, $H(X|\hat{X}_{2},U)$
and $H(X|U)$ .

\section*{Appendix D: Sketch of Proof of Achievability for Proposition \ref{prop:act_stat}}

\begin{figure}[h!]
\centering\includegraphics[bb=37bp 242bp 570bp 678bp,clip,scale=0.6]{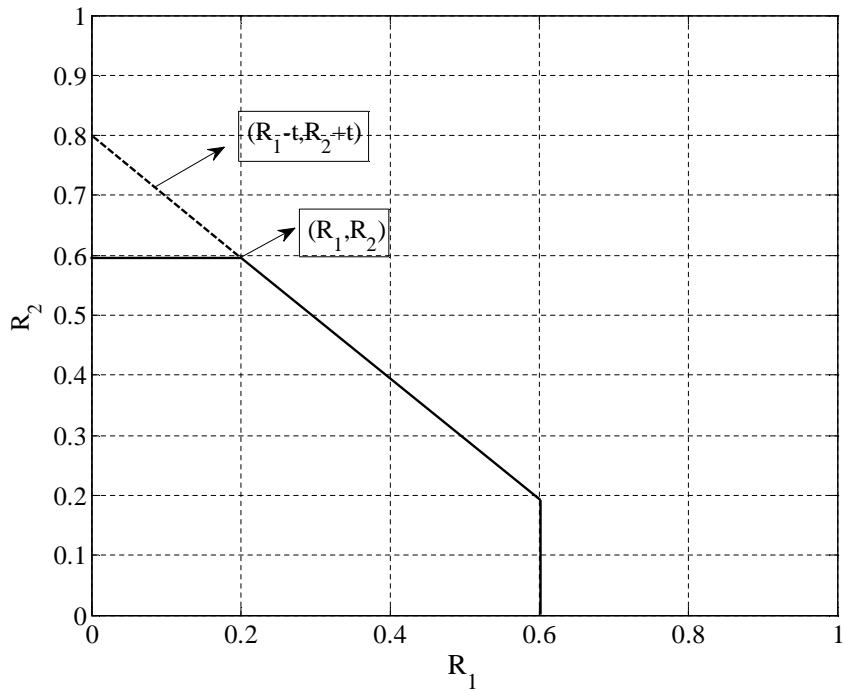}
\caption{Illustration of the rate regions (\ref{eqn: channel_region}) (dashed
lines) and (\ref{eqn: channel_region_equiv}) (solid lines).}

\label{fig:equivalence} 
\end{figure}

We will prove below that the following rate region is achievable \begin{subequations}\label{eqn: channel_region_equiv}
\begin{eqnarray}
R_{1} & \leq & H(\ce{f}(A))\label{eq:R1_equiv}\\
R_{1}+R_{2} & \leq & I(A,U;Y)-I(U;S|A),\label{eq:R1+R2_equiv1}\\
R_{2} & \leq & I(A;Y|\ce{f}(A))+I(A,U;Y|A)-I(U;S|A),\label{eq:R1+R2_equiv2}
\end{eqnarray}
 \end{subequations} for a given joint distribution as in (\ref{eq:joint_channel}).
Assuming now that this rate region is achievable, we show that the
rate region (\ref{eqn: channel_region}) is also achievable. Region
(\ref{eqn: channel_region}) is larger than (\ref{eqn: channel_region_equiv})
owing to the absence of the inequality (\ref{eq:R1+R2_equiv2}). The
two regions are illustrated in Fig. \ref{fig:equivalence} for a given
choice of the distribution 
(\ref{eq:joint_channel}), with region (\ref{eqn: channel_region_equiv})
in solid lines and (\ref{eqn: channel_region}) in dashed lines. We
now argue that the achievability of region (\ref{eqn: channel_region_equiv})
(solid lines) implies the achievability of region (\ref{eqn: channel_region})
(dashed lines) as well, by following the same arguments as in \cite{KornerMarton}.
Specifically, we observe that, if $(R_{1},R_{2})$ is achievable with
some scheme, then $(R_{1}-t,R_{2}+t)$ is also achievable for all
$0\le t\le R_{1}$. This is due to the fact that, if the rate pair
$(R_{1},R_{2})$ is achievable, then some of the rate of the common
message $M_{1}$ can always be transferred to the private message
$M_{2}$ for Decoder 2 to achieve $(R_{1}-t,R_{2}+t)$ if $0\le t\le R_{1}$.
It follows immediately that all the points on the dashed line in Fig.
\ref{fig:equivalence} are also achievable.

The discussion above allows us to conclude that concludes that, if
region (\ref{eqn: channel_region_equiv}) is achievable, then the
desired rate region (\ref{eqn: channel_region}) is also achievable.
We now focus on proving the achievability of (\ref{eqn: channel_region_equiv}).
To this end, we combine superposition coding and the technique proposed
in \cite{Weissman}. Fix the joint distribution as in (\ref{eq:joint_channel}).
We first generate the codebook $b^{n}(m_{1})$, $m_{1}\in[1:2^{nR_{1}}]$,
i.i.d. with pmf $p(b)$. Next, we generate a superimposed codebook
for each $b^{n}$ of $a^{n}(m_{1},m_{2})$ codewords, $m_{2}\in[1:2^{nR_{2}}]$,
i.i.d. with pmf $p(a|b)$. For every $a^{n}$ sequence, a codebook
of $u^{n}(m_{1},m_{2},j)$ sequences is generated, $j\in[1:2^{n\tilde{R}}]$,
i.i.d. with pmf $p(u|a)$.

To encode messages $(m_{1},m_{2})$, Encoder selects the codeword
$a^{n}(m_{1},m_{2})$, and chooses a $u^{n}$ codeword jointly typical
with action and state sequence, which requires $\tilde{R}\ge I(U;S|A)$.
Then $x_{i}=g(u_{i},s_{i})$ is then sent through the channel. Decoder
1 decodes the message $m_{1}$ correctly if $R_{1}\le H(B)$. Decoder
2 looks for the unique pair of messages $(m_{1},m_{2})$ such that
the tuple $(y^{n},b^{n}(m_{1}),a^{n}(m_{1},m_{2}),u^{n}(m_{1},m_{2},j))$
is jointly typical for some $j\in[1:2^{n\tilde{R}}]$. This step is
reliable if $R_{1}+R_{2}+\tilde{R}\le I(A,U;Y)$ and $R_{2}+\tilde{R}\le I(U,A;Y|B)=I(A;Y|B)+I(U;Y|A)$.
Using Fourier-Motzkin elimination to eliminate rate $\tilde{R}$ leads
to the bounds (\ref{eqn: channel_region_equiv}).

\end{document}